\documentclass[english,aps,floats,twocolumn,epsf,prb,showpacs,floatfix,nofootinbib,amsmath,amssymb]{revtex4-2}
\usepackage[utf8]{inputenc} 
\usepackage[T1]{fontenc} 
\usepackage{amsmath, amssymb}
\usepackage{soul}

\usepackage{graphicx}
\usepackage[caption=false]{subfig}
\usepackage{epstopdf}
\usepackage{hyperref}
\hypersetup{colorlinks=true,urlcolor=blue,citecolor=blue,linkcolor=blue,breaklinks=true}
\usepackage[capitalise]{cleveref}
\usepackage{color}
\usepackage{braket}
\usepackage{float}
\usepackage{microtype}
\usepackage{lipsum}
\usepackage{blindtext}
\usepackage{layouts}
\usepackage[normalem]{ulem}

\definecolor{darkred}{RGB}{165,50,50}
\definecolor{darkgreen}{RGB}{0,100,0}

\newcommand{\CW}[1]{{\color{blue} {#1}}}
\newcommand{\cw}[1]{{\color{blue}[\textbf{CW:} #1]}}
\newcommand{\lr}[1]{{\color{darkgreen}[\textbf{LR:} #1]}}
\newcommand{\pw}[1]{{\color{red}[\textbf{PW:} #1]}}

\newcommand{\kh}[1]{{\color{red}[\textbf{KH} #1]}}
\definecolor{lightgray}{RGB}{211,211,211}
\newcommand{\old}[1]{{\color{lightgray}{\sout{#1}}}}

\renewcommand{\CW}[1]{}
\renewcommand{\cw}[1]{}
\renewcommand{\lr}[1]{}
\renewcommand{\pw}[1]{}
\renewcommand{\kh}[1]{}
\renewcommand{\old}[1]{}

\newcommand{\TMPWW}{86mm}
\newcommand{\me}{\mathrm{e}}
\newcommand{\md}{\mathrm{d}}
\newcommand{\mi}{\mathrm{i}}
\newcommand{\operator}[1]{\hat{#1}}
\newcommand{\creation}[1]{\operator{c}^\dagger_{#1}}
\newcommand{\annihilation}[1]{\operator{c}^{\phantom\dagger}_{#1}}

\newcommand{\Tr}{\mathrm{Tr\,}}

\newcommand\norm[1]{\left\lVert#1\right\rVert}

\newenvironment{psmallmatrix}
{\left(\begin{smallmatrix}}
	{\end{smallmatrix}\right)}

\begin{document}

\title{Photoexcitations in the Hubbard model -- 
generalized Loschmidt amplitude analysis of impact ionization in small clusters}

\author{C. Watzenb\"ock, M. Wallerberger, L. Ruzicka, P. Worm, K. Held and  A. Kauch}

\affiliation{Institute of Solid State Physics, TU Wien, 1040
Vienna, Austria}

\date{ \today }

\begin{abstract}
%
%
We study photoexcitations in small Hubbard clusters of up to 12 sites. After the electric field pulse some of these clusters show an increase of the double occupation through impact ionization. We treat the time-dependent electromagnetic field classically and  calculate time evolution  by exact diagonalization. As a tool for better analyzing the out-of-equilibrium dynamics, we generalize the Loschmidt amplitude. This way, we are able to resolve which many-body energy eigenstates are responsible for impact ionization and which ones show pronounced changes in the double occupation and spin energy. Our analysis reveals that the increase of spin energy is of little importance for impact ionization.
We further demonstrate that, for one-dimensional chains, the optical conductivity has a characteristic peak structure originating solely from vertex corrections. 
\end{abstract}

\maketitle


\section{Introduction}
\label{sec:IntroLoschmidtSpin}
Light-induced phenomena in strongly correlated systems have gained much attention recently, not only because of the advance in pump-probe laser experiments \cite{Giannetti2016,Zonno21} but also for solar energy conversion \cite{manousakis2010,Assmann2013,Wang2015,werner2014,sorantin2018,Manousakis2019,Holleman2016,Kauch2020Disorder,Maislinger2020}. The particular advantage of a strong electron-electron interaction  for solar energy conversion is impact ionisation~\cite{manousakis2010,werner2014,sorantin2018,Manousakis2019,Holleman2016,Wais2020,Kauch2020Disorder,Maislinger2020}, which allows for the generation of multiple electron-hole pairs (aka doublons and holons) per photon. This is one way
to boost the efficiency of
solar cells beyond the Shockley-Queisser limit~\cite{Shockley1961}
of 30-34\%.
%
Impact ionization is a genuine nonequilibrium process, which is particularly challenging to describe in theory
if electronic correlations are strong, as weak-coupling perturbation theory \cite{manousakis2010} or the Boltzmann equation \cite{Wais2020} cannot be reliably applied. One possibility is to employ nonequilibrium  dynamical mean-field theory \cite{Aoki2014,werner2014,sorantin2018} which treats local correlations non-perturbatively, another route is to study the time evolution directly for small clusters \cite{Alvermann2011,Innerberger2020,Maislinger2020,Kauch2020Disorder}. In Refs.~\onlinecite{Maislinger2020,Kauch2020Disorder} it was shown by studying nonequilibrium spectral functions and the double occupation that the effect of impact ionization can occur even in small clusters. It manifests itself in the rise of the double occupancy after the light pulse is turned off. This effect was found to be enhanced by disorder and next-nearest neighbor hopping~\cite{Kauch2020Disorder}, but strongly suppressed in the case of a simple one-dimensional chain geometry.

As a tool to study such nonequilibrium dynamics, we use  the Loschmidt amplitude:
\begin{equation}
L^{\ket{\psi}}(\tau) \equiv \bra{\psi} \me^{- \mi \tau \hat H} \ket{\psi}.
\label{eqn::L_def1}
\end{equation}
Its Fourier transform is a decomposition of the wave function $\ket{\psi}$ with respect to the eigenstates of the Hamiltonian $\hat{H}$.
The module squared of the Loschmidt amplitude, called Loschmidt echo or fidelity~\cite{Gorin2006,brush1966kinetic},
is used as a measure of time irreversibility and was measured, as early as 1950, in NMR experiments~\cite{Hahn1950}. It has recently gained popularity also in the field of dynamical quantum phase transitions~\cite{Heyl2013}, since  nonanalyticities in its logarithm correspond to a generalized phase transition in time  \cite{Heyl2019}. 
The Loschmidt amplitude
and the related work distribution function have also been used for studying quantum quenches~\cite{Andrei2019,Palmai2014} and impact ionization~\cite{Maislinger2020}.
   
The Loschmidt amplitude~(\ref{eqn::L_def1}) allows us to study the non-equilibrium dynamics with respect to energy eigenstates only. In an effort to understand the behavior of the double occupation, in this work we generalize the Loschmidt amplitude, allowing us to simultaneously resolve dynamics of multiple quantities, such as energy and double occupation.
We further show that there is a clear relation between the Loschmidt amplitude and the nonequilibrium Green's function as well as to the optical conductivity.
As an application of the generalized Loschmidt amplitude, we study the dynamics and redistribution of  the double occupation and Heisenberg spin energy  and its correlation to the many-body energy eigenvalues. We focus on times  after the electric field pulse has been switched off. Then, impact ionization occurs for some of the  $12$-site Hubbard clusters but not for others. Surprisingly, the observed impact ionization happens predominately when already at least two doubly occupied sites are present. 

In our earlier study, Ref.~\onlinecite{Kauch2020Disorder}, impact ionization in several $12$-site systems was analyzed by studying the double occupancy and spectral functions after an electric field pulse. It was conjectured that spin fluctuations compete with impact ionization, which might be an explanation why in one-dimensional chains with only nearest-neighbor hopping no impact ionization was found. We find that at least for the strong electric field strength considered, the spin fluctuations do not compete with and thus do not suppress impact ionization.
In one-dimensional chains more of the initial spin-spin correlation survives, at the same time the change in Heisenberg spin energy after the pulse is smaller than for geometries with larger connectivity.
A further difference between the aforementioned one-dimensional chains and other geometries is that for the chains vertex corrections to the optical conductivity dominate and lead to sharp absorption peaks.

The paper is organized as follows: 
\Cref{sec:Methods} outlines the used model and gives details on the time-evolution algorithm.
In \cref{sec:FGR_and_L},  we introduce the Loschmidt amplitude and our novel generalization of it. This is supplemented in \cref{App:Properties}, where its properties and relation to a probability distribution are given.  
In \cref{sec:connection} we present the connection of the Loschmidt amplitude to other physical quantities. We show how light absorption can be analyzed with the Loschmidt amplitude and we elaborate on the relation to Fermi's golden rule. Further, we show how the Loschmidt amplitude is expressed in the quantum many-body picture and can be partly represented by means of simple one-particle excitations with the Green's function formalism. 
We further highlight which features are solely due to vertex corrections and thus cannot be captured by one-particle Green's functions.
In \cref{sec:GL_all}, we show the results for our generalization of the Loschmidt amplitude and determine which energy states are responsible for the long-term dynamics. In \cref{sec:GL_docc}, we apply the generalized Loschmidt amplitude to the double occupancy to elucidate the phenomenon of impact ionization. 
In \cref{sec:SpinExcitations} we show the spin correlation function and the different energy scales to determine the importance of spin excitations in $12$-site systems. 
We finally apply the generalized Loschmidt amplitude to a measure of spin-correlations in \cref{sec:GL_spin} and summarize our findings in \cref{sec:Conclusions}.

\section{Hubbard Model and Time Evolution}
\label{sec:Methods}

\subsection{Hamiltonian} As a prototypical Hamiltonian for strongly correlated electrons systems we consider the Hubbard model~\cite{hubbard1963}:  
\begin{equation}
\operator{H} = \sum_{i,j,\sigma} v_{ij} \creation{j\sigma}\annihilation{i\sigma} + U\sum_{i} \hat n_{i\uparrow} \hat n_{i\downarrow}
\label{eq:Hubbard}.
\end{equation}
Here, $\creation{i\sigma}$ ($\annihilation{i\sigma}$) creates (annihilates) an electron on site $i$ with spin $\sigma$, $\hat n_{i\sigma}=\creation{i\sigma}\annihilation{i\sigma}$ is the occupation number operator,  $U$ is the  (screened) Coulomb interaction between electrons on the same site, and $v_{ij}$ for $i\neq j$ describes the hopping amplitude from site $i$ to $j$. 
In the following, we assume the system to be half-filled, meaning on average each site has $0.5$ spin-up electrons and $0.5$ spin-down electrons. 

\subsection{Light pulse}
\label{sec:LightPulseAndUnits}
The interaction with light is described by coupling an external, classical electric field pulse $\vec E(t)$ to the system by Peierls substitution~\cite{Peierls1933} in a gauge where $\vec E(t) = - \partial_t \vec{A}(t)$ (sometimes referred to as Weyl gauge). 
Under Peierls substitution the time dependence enters only as a phase factor in the hopping terms
\begin{equation}
v_{ij} \rightarrow v_{ij} \me^{\mi q \chi_{ij}(t)} = v_{ij} \me^{-\mi q \int_{R_i}^{R_j} \md  \vec{r^\prime} \: \vec{A}(\vec{r^\prime}, t)}.
\label{eqn::Peierls}
\end{equation}
Here, we are working in natural units where the charge of the electron $q=1$, Plank's reduced constant $\hbar=1$ and the geometric lattice spacing $\tilde{a}=1$. 
All energies in this paper are in units of the nearest neighbor hopping term $|v_{ij}|\equiv1$. This implies a unit of time of $\hbar/|v_{ij}|$. Typical hopping values in correlated systems are around $0.5$eV, which corresponds to time units of $\sim1$ fs.     

\begin{figure}
	\centering
	\includegraphics[width=0.9\linewidth]{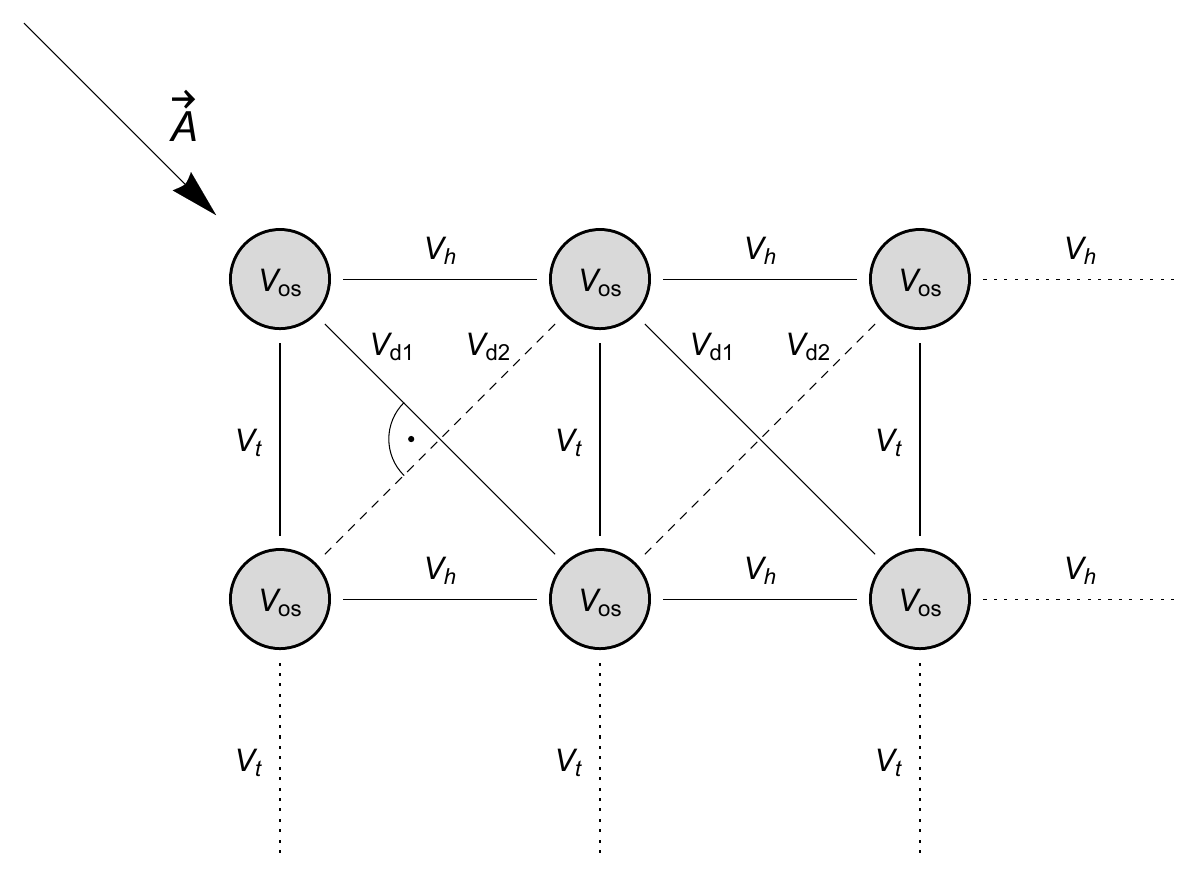}
	\caption{Example of a  $2\times 3$ box geometry (possibly further extended) with  on-site potential equal on all sites $v_{ii}=v_{os}$, NN hopping $v_h$ in horizontal direction and $v_t$ in the vertical direction as well as two different NNN (diagonal) hoppings $v_{d1}$ and  $v_{d2}$. In our simulations, the time-independent prefactors are equal for NN hoppings $v_t=v_h\equiv 1$ and  for NNN hoppings $v_{d1}=v_{d2}\equiv v_d$ .
		The vector potential $\vec{A}$ is chosen along one of the diagonal directions (as shown in the Figure and employed in our calculation), the parameter $a$ describing the strength of the field is the same for $v_h$ and $v_t$, it is twice as big for $v_{d1}$, and zero for $v_{d2}$.}
	\label{fig:geometry}
\end{figure}

We also assume that the wavelength of light is much larger than the system size (optical light). Therefore we use a vector potential independent of $\vec{r}$.
For modeling of the $\vec E-$field we follow Ref.~\onlinecite{Kauch2020Disorder} and  Ref.~\onlinecite{Maislinger2020} and choose 
\begin{equation}
\vec A(t) = \frac{\vec E_0}{\omega_p} \left[\cos(\omega_p(t-t_p))-\cos(\omega_p t_p)  \right] \, \me^{-\frac{(t-t_p)^2}{2\sigma_p^2}}, 
\label{eq:EM_pulse}
\end{equation}
which approximately corresponds to an $\vec E$-field of 
\begin{equation}
\vec E(t) = \vec E_0 \sin(\omega_p(t-t_p)) \,  e^{-\frac{(t-t_p)^2}{2\sigma_p^2}}
\end{equation}
for $1/\omega_p \ll \sigma_p$. 
For all the simulations, we applied the field in-plane and under a $45 ^{\circ}$ angle, as illustrated in Fig.~\ref{fig:geometry}. 
For convenience we combined the magnitude of the $E-$field together with the lattice constant $\tilde a$, and the frequency of the pulse $\omega_p$ into a single, directional-dependent dimensionless parameter $a \equiv q \vec E_0 \cdot \vec e_x  \tilde a/(\hbar \omega_p) $, where $\vec{e_x}$ is the unit vector in $x$-direction.

\subsection{Geometries and model parameters of the systems considered}
\label{sec::parameters}

In this work we mainly focus on $12$-site systems ($N_s=12$) of three different geometries: $4\!\times\!3$ and $6\!\times\!2$ boxes and a $12\!\times\!1$ chain. We use open boundary conditions (OBC), i.e.,  no hopping is possiblepossible from the leftmost sites to the leftleft, and from the rightmost sites to the rightright.  
Additionally, we also considered a $12\!\times\!1$ system with periodic boundary conditions (PBC). We implemented the $A$-field for PBC in such a way that it does not break the translation symmetry. Strictly speaking, this idealization would not correspond to an electric field but rather a magnetic field through a closed ring that entails a circular E-field along the ring. The magnetic field does not couple to the spin. 

The interaction for the $12$-site systems is always set to $U=8$. For the frequency and field strength we chose, to allow for direct comparisons, the same parameters as in \cite{Kauch2020Disorder}, i.e., $\omega_p=11$, $\sigma_p=2$ and $a=0.8$, unless specified otherwise. These parameters are close to optimal for obtaining the strongest impact ionization (see \cref{App:Scan} for a parameter scan).

We also present selected results for an $8\times 1$ system, for which we 
were still able to fully diagonalize the Hamiltonian. 


\subsection{Time evolution} 
In order to calculate the time evolution of the system driven out of equilibrium by a time-dependent light pulse, we solve the time-dependent Schr\"odinger equation
\begin{equation}
\label{eq:schroedinger}
\mi \partial_t \ket{\psi(t)} = \hat{H}(t) \ket{\psi(t)},
\qquad \ket{\psi(0)}=\ket{\psi_0}
\end{equation}
using optimized commutator-free Magnus integrators of fourth order described in detail as \texttt{CF4oH} in Ref.~\cite{auzinger2021}. As an error tolerance for the adaptive time-stepping algorithm we used $\texttt{tol} < 10^{-6}$. We use the Lanczos method with an on-the-fly representation of $\hat{H}$~\cite{Wallerberger2022}.
The tolerance for the adaptive Lanczos method for the matrix exponential was set to less than $10^{-8}$.
The initial state $\ket{\psi_0}$ of the time evolution is always taken to be the (non-degenerate) ground state.

\section{Loschmidt amplitude and generalized Loschmidt amplitude}
\label{sec:FGR_and_L}

\subsection{Loschmidt amplitude}

A one-operator Loschmidt amplitude may be defined for a hermitian operator $\hat A = \hat A^\dagger$ as 
\begin{equation}
L_A^{\ket \psi}(\bar \alpha) = \bra{\psi} \me^{-\mi \bar \alpha \hat A}  \ket{\psi}.
\end{equation}
From this, it directly follows that
\begin{equation}
    L_A^{\ket \psi}(\bar \alpha)^* = L_A^{\ket \psi}(- \bar \alpha)
\end{equation}
and that its Fourier transform is given by:
\begin{equation}
\begin{split}
L_A^{\ket \psi}(\alpha)
&=\int \md \bar{\alpha} \:  \me^{\mi \bar{\alpha} \alpha}L_A^{\ket \psi}(\bar \alpha) \\
&= 2\pi \sum_n |\langle\psi|a_n\rangle|^2 \delta(\alpha - \alpha_n),
\end{split}
\label{eqn::L_def2a}
\end{equation}
where the spectral representation of $\hat A$ is $\hat A = \sum_n \alpha_n|a_n\rangle\langle a_n|$.
Hence, $\frac{1}{2\pi} L_A^{\ket{\psi}}(\alpha)$ has the properties of a probability density function.
For $\hat A = \hat H$ and $\bar \alpha=\tau$ (time), we recover \cref{eqn::L_def1}.
The Fourier transform~(\ref{eqn::L_def2a}) of \cref{eqn::L_def1} is then
the energy spectrum projected on a given state $\ket{\psi}$~%
\cite{Maislinger2020, Kennes2021Loschmidt}.  

 In this work, the Hamiltonian~\eqref{eq:Hubbard} is time dependent due to 
 the time-dependent electromagnetic field, cf.~\cref{eqn::Peierls}. Here and in the following, we will use the Loschmidt amplitude w.r.t. the Hamiltonian at $t_0=0$, $\hat H\equiv\hat H(t_0=0)$, before the onset of the electromagnetic pulse.  The Fourier transform~(\ref{eqn::L_def2a})
 then reads:
\begin{equation}
L(t,\omega) \equiv L^{\ket{\psi(t)}}( \omega ) 
   =  2 \pi \sum_n \left|  \braket{\psi(t)|E_n}  \right|^2 \delta(\omega - E_n),
\label{eqn::L_def2}
\end{equation}
where $\hat H(0)|E_n\rangle=E_n|E_n\rangle$ and $|\psi(t)\rangle$ is the
time dependent wave function, the time evolution of which is governed by \cref{eq:schroedinger}.
In the following, we set the ground-state energy to zero, as it simplifies some subsequent equations and leads to the property that $L^{\ket{\psi}}(\omega < 0) = 0$.

Note that $|E_n\rangle$ satisfy the Schr\"odinger equation~\eqref{eq:schroedinger} for  $t\to\pm\infty$ only.  This means that $L(t,\omega)$ is
the projection of the wave function at finite time $t$ onto asymptotic ``scattering channels''  $|E_n\rangle$ with energy $\omega=E_n$.  Similarly, the correspondence of the Loschmidt parameter $\tau$ to time $t$ is only valid asymptotically:
\begin{equation}
    \lim_{t\to\pm\infty} U(t + \tau, t) \propto \me^{-\mi\tau \hat H(0)},
\end{equation}
where $U(t,t')$ is the full time evolution operator.  $L(t,\omega)$ thus allows us to track the redistribution of spectral weight between different asymptotic states with energy $\omega$ as a function of time $t$, mapping out the effect of the light pulse on the system prepared in some initial state $\ket{\psi_0}$.


The major advantage of using the Loschmidt amplitude for describing the system is that it can be efficiently computed, e.g.,  through time-propagation with commutator-free Magnus integrators (or density matrix renormalization group for one-dimensional systems \cite{Kennes2021Loschmidt}) even if the dimension of the Hamiltonian is so large that it prevents direct diagonalization.
After the light-pulse fades away, i.e., for $t\gg t_p$, the Loschmidt amplitude is independent of time. 
For cases where all the eigenstates $\ket{E_n}$ of the given operator are known, one can directly use Eq.~\eqref{eqn::L_def2} to obtain the Loschmidt amplitude. However, for a general operator like the Hamiltonian, where the eigenstates are a priori unknown, it is advantageous to work directly with Eq.~\eqref{eqn::L_def1}.

\subsection{Generalized Loschmidt amplitude}

In the previous subsection, the Loschmidt amplitude $L_{\hat A}^{\ket{\psi}}$ was interpreted as a spectral decomposition of the state $\ket{\psi}$ with respect to the eigenstates of $\hat A$. 
We now want to generalize this concept and decompose the state with respect to two operators at the same time. This corresponds to the joint probability distribution of $\hat A$ and $\hat B$, where $\hat A^\dagger = \hat A$ and $\hat B^\dagger = \hat B$. It reads 
%
\begin{equation}
L_{AB}^{\ket{\psi}}(\bar \alpha, \bar \beta ) \equiv \bra{\psi} \me^{- \mi \bar \alpha \hat A} \me^{- \mi \bar \beta \hat B} \ket{\psi} . 
\label{eqn::LAB_def1}
\end{equation} 
The Fourier transform of \cref{eqn::LAB_def1} with respect to $\bar \alpha \rightarrow \alpha$  and $\bar \beta \rightarrow \beta$ is given by
\begin{equation}
\begin{array}{rcll}
L_{AB}^{\ket{\psi}}( \alpha, \beta ) &=&  (2 \pi)^2 \sum_a \sum_b 
&\braket{\psi | a} \braket{ a | b}  \braket{b | \psi} 
\\&&&\times\delta(\beta - b) \, \delta(\alpha - a),
\end{array}
\label{eqn::LAB_def2}
\end{equation} 
where $\hat A \ket a = a \ket a$ and $\hat B \ket b = b \ket b$. An extension to more than two operators is straightforward, though at present computationally not feasible. 
If the operators $\hat A$, $\hat B$ do not commute,  $L_{AB}^{\ket{\psi}}( \alpha, \beta )$ is a complex quantity. For the interpretation as a probability distribution the real part is sufficient (see \cref{App:Properties}).



\section{Connection of the Loschmidt amplitude to other physical quantities}
\label{sec:connection}

\subsection{Visualizing Fermi's golden rule with Loschmidt amplitude}
Using the Loschmidt amplitude, we can identify which eigenstates of the unperturbed Hamiltonian are excited by the perturbation through the $A$-field (see also Ref.~\onlinecite{Maislinger2020}). For small perturbations,
one may resort to a Fermi's golden rule (FGR) description.
We expect FGR to hold for short times before the pulse in Eq.~\eqref{eq:EM_pulse} reaches its full strength.

To make a connection with FGR, we split the Hamiltonian into a static part and a time-dependent perturbation. The static part is given by Eq.~\eqref{eq:Hubbard} with time-independent $v_{ij}$.
The rest constitutes the dynamic part and may be expanded with respect to $\vec A$ as 
\begin{equation}
\begin{array}{rcl}
\hat H_{\mathrm{dyn}}(t) &=& \sum_{ij\sigma} \left( \me^{\mi \chi_{ij}(t)} - 1\right) v_{ij} \hat c^\dagger_{i \sigma} c_{j \sigma} \\
&=&  \vec{\hat J} \cdot \vec{A}(t) + \mathcal O (\vec{A}^2)
\end{array}
\label{eqn::Hdyn}
\end{equation} 
with the current operator 
\begin{equation}
\vec{\hat J} = - \mi \sum_\sigma \sum_{ij} \left( \vec{R}_i - \vec{R}_j \right) v_{ij} \, \hat c^\dagger_{i \sigma} c_{j \sigma}.
\label{eqn::CurrentOp}
\end{equation}
For convenience we define the projection of the current $ \vec{\hat J}$ onto the direction of the $\vec{A}$ field as $\vec{\hat J} \vec A(t) \equiv \hat j f(t) $ with $f(t)= a \left[\cos(\omega_p(t-t_p))-\cos(\omega_p t_p)  \right] \me^{-\frac{(t-t_p)^2}{2\sigma^2}}$.  

For a large pulse width $\sigma$ one ends up with a perturbation $\propto \cos(\omega_p t) $ and one can directly apply the textbook version of Fermi's golden rule according to which the probability amplitude of the transition from an initial state $\ket{i}$ to a final state $\ket{f}$ is given by
\begin{equation}
\Gamma_{i \rightarrow f} = 2 \pi \left| \bra{f} \hat j \ket{i}  \right|^2 \mathcal{N}(\omega),
\label{eqn::FGR}
\end{equation}
where $\mathcal{N}(\omega)=\frac{1}{\mathrm{dim}  H } \sum_n \delta(\omega - E_n)$ is the \textit{ density of states} of the system and $E_n$ are the many-body eigenenergies 
\footnote{The usual way to calculate the density of states would require a full diagonalization of the Hamiltonian and is thus only feasible for rather small system sizes. For the $8\!\times\!1$ system $\mathrm{dim} {H} = \begin{psmallmatrix} 8 \\ 4 \end{psmallmatrix}^2 = 4900$ this was possible and is shown in \cref{fig:Loschmidt_8x1}. For the various $N_s=12$ systems where $\mathrm{dim} {H} = \begin{psmallmatrix} 12 \\ 6 \end{psmallmatrix}^2 = 853776$ the calculation of the density of states was unfeasible due to memory constraints.}.
Please note that this version of FGR requires a long time $t\gg 1/\omega_p$,  but at the same time, a weak perturbation which for our $A$-field strength translates in a not too long time.

A simple way of visualizing which final states are allowed within the first-order perturbation theory (FGR) is by means of the Fourier transform of the Loschmidt amplitude with respect to  $\hat j \ket{\psi_0}$:
\begin{equation}
L^{\vec{\hat{j}} \ket{\psi_0}}(\tau) =  \bra{\psi_0} \hat j \;  \me^{- \mi \tau \hat H} \; \hat j \ket{\psi_0}.
\label{eqn::L_FGR1}
\end{equation}
We will refer to this function as the Loschmidt amplitude for optical absorption for reasons that will become clear in \cref{sec:connection_GF_and_L}. The Fourier transform reads
\begin{equation}
L^{\hat j \ket{\psi_0}}(\omega) =  2 \pi \sum_n \left| \bra{E_n} \hat j \ket{\psi_0} \right|^2 \delta(\omega - E_n),
\label{eqn::L_FGR2}
\end{equation}
and is just given by a sum over the possible final states of Eq.~\eqref{eqn::FGR}. 

\subsubsection{\texorpdfstring{$8\!\times\!1$}{8 x 1} system}

\begin{figure}[htb]
	\centering
	\includegraphics[width=\linewidth]{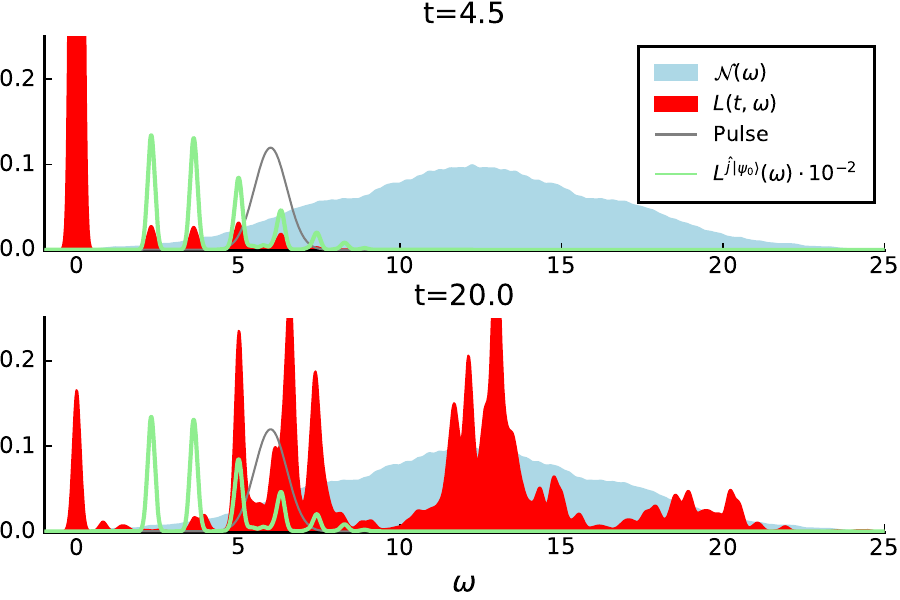}
	\caption{ Density of states $\mathcal{N}(\omega)$ and Loschmidt amplitude $L(t,\omega)\equiv L^{|\psi(t)\rangle}(\omega)$ for a $8\times1$ geometry with $U=4$ NN-hopping only. The light-green line ($L^{\hat j \ket{\psi_0}}$) shows the allowed transitions by first order perturbation theory from the ground state. The gray curve ("Pulse") shows the Fourier-transform of the $A$-field in arbitrary units, which has a base frequency $\omega_p=6$ 
		maximal amplitude at time $t_p=8$  and pulse width $\sigma_p=2$. Peaks are broadened with $\sigma_\omega = 0.09$, meaning that a delta-peak in frequency is depicted as a Gaussian with standard-deviation $\sigma_\omega$.}
	\label{fig:Loschmidt_8x1}
\end{figure}

In Fig.~\ref{fig:Loschmidt_8x1}, we show the Loschmidt amplitude \eqref{eqn::L_def2}, $L^{|\psi(t)\rangle}(\omega)$, as well as the FGR-allowed transitions given by $L^{\hat j \ket{\psi_0}}(\omega)$ of Eq.~\eqref{eqn::L_FGR1} for an $8\!\times\!1$ system at two different times: $t=4.5$ (at the onset of the EM pulse that is centered at $t_p=8$) and $t=20$.

At short times (or small electric fields), only transitions allowed by $L^{\hat j \ket{\psi_0}}(\omega)$ are possible, which is evident by comparing  $L^{|\psi(t)\rangle}(\omega)$ with $L^{\hat j \ket{\psi_0}}(\omega)$ in the top panel of Fig.~\ref{fig:Loschmidt_8x1}.
The maximal absorption rates can only be achieved when the pulse frequency $\omega_p$ has a large overlap, or resonance, with the allowed states. 
The Fourier transform of the electric field (in arbitrary units) is also shown in Fig.~\ref{fig:Loschmidt_8x1} as a gray curve centered around $\omega_p=6$ with $\sigma_p=2$. 
For the allowed transitions, we find a distinct peak structure. A more detailed description of the peaks can be found later in \cref{sec:connection_GF_and_L}.
Also the many-body density of states $\mathcal{N}(\omega)$ is shown with a broad distribution of eigenstates. This 
is due 
to the rather weak interaction of $U=4$.  
For large interaction values, closer to the atomic limit, the density of states will only have contributions around $Un$ with $n \in \mathbb{N}_0$, where each contribution can be associated with a specific double occupation value.

The Loschmidt amplitude $L^{|\psi(t)\rangle}(\omega)$ for a later time, $t=20$ can no longer be described by the FGR (see lower panel of Fig.~\ref{fig:Loschmidt_8x1}). When the energy levels allowed by $L^{\hat j \ket{\psi_0}}(\omega)$ are at least partially occupied the pulse can excite the system further. 
If higher energy states are available and allowed by the selection rules, the predominant excitations are around integer multiples of the pulse frequency $\omega_p$. One may refer to these as single-, double- and triple-photon excitations, although the electric field was treated as classical. The multi-photon excitations in this sense are, however, found to be retarded with respect to the single-photon excitations, signifying sequential absorption (see also \cref{fig::FGRandLs}(b)-(f)).

\subsubsection{12-site systems}

Motivated by the study of Refs.~\cite{Maislinger2020,Kauch2020Disorder}, where impact ionization was found to occur in $4\!\times\!3$ and $6\!\times\!2$  clusters, but not in the $12\!\times\!1$ chains, we analyzed these systems in more detail. For the $12\!\times\!1$ systems, we considered both open and periodic boundary conditions (OBC and PBC). For the chains with OBC we considered systems with nearest-neighbor (NN) hopping only as well as a frustrated system with next-nearest-neighbour (NNN) hopping $v'=0.5$. These systems were chosen to investigate the role of spin frustration, which we discuss later in \cref{sec:SpinExcitations,sec:GL_spin}.

\begin{figure*}[t]
	\centering
	\subfloat[Allowed transitions by 1. order PT]{
		\includegraphics[width=\TMPWW, 
		trim=0cm 4mm 0cm 0cm, clip=true]{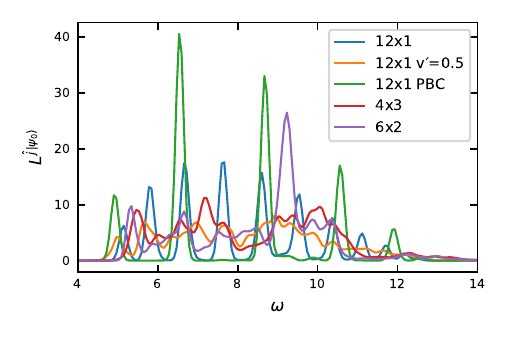}
	}
	\subfloat[$12\!\times\!1$ $v'=0.0$]{
		\includegraphics[width=\TMPWW, 
		trim=0cm 4mm 0cm 0cm, clip=true]{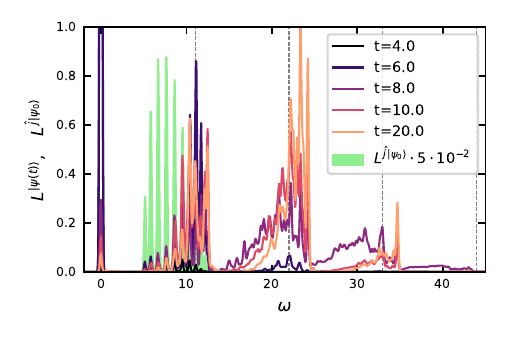}
	}
	\\[-3mm]
	\hspace{0mm}
	\subfloat[$12\!\times\!1$ $v'=0.5$]{
		\includegraphics[width=\TMPWW,
		trim=0cm 4mm 0cm 0cm, clip=true]{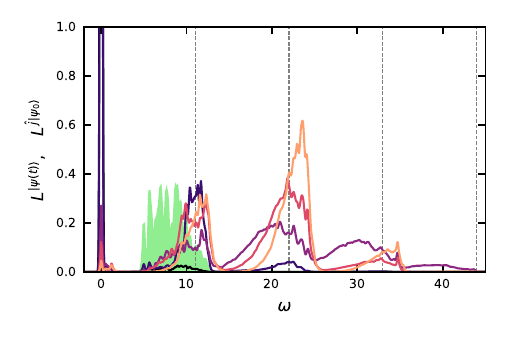}
	}
	\subfloat[$12\!\times\!1$ PBC]{
		\includegraphics[width=\TMPWW,
		trim=0cm 4mm 0cm 0cm, clip=true]{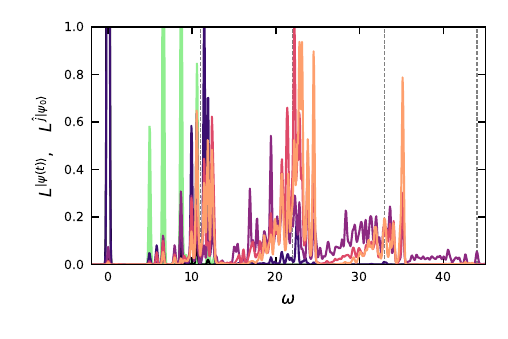}
	}
	\\[-3mm]
	\hspace{0mm}
	\subfloat[$4\!\times\!3$]{
		\includegraphics[width=\TMPWW, 
		trim=0cm 4mm 0cm 0cm, clip=true]{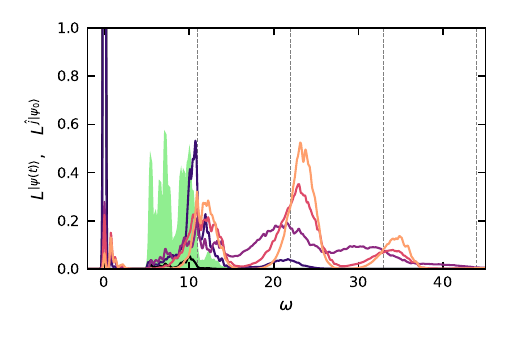}
	}
	\subfloat[$6\!\times\!2$]{
		\includegraphics[width=\TMPWW, 
		trim=0cm 4mm 0cm 0cm, clip=true]{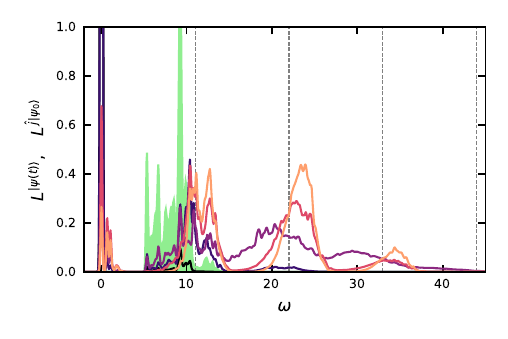}
	}
	\\[-3mm]
	\caption{Allowed transitions by first order perturbation theory $L^{\hat j \ket{\psi_0}}(\omega)$ and time resolved Loschmidt amplitude $L^{|\psi(t)\rangle}(\omega)$ for different $N_s$=12 systems. In (a) only $L^{\hat j \ket{\psi_0}}(\omega)$ is shown for all systems. In (b)-(f) $L^{|\psi(t)\rangle}(\omega)$ is shown for the five  $N_s$=12 systems at several times during the pulse (centered at $t_p=8$) and after it. The dashed gray lines mark integer multiples of the pulse frequency $\omega_p$. After $t=20$ $L^{|\psi(t)\rangle}(\omega)$ does not change any more. The legend for (c)-(f) is the same as for (b).}
	\label{fig::FGRandLs}
\end{figure*}

In \cref{fig::FGRandLs}(a) we show the FGR-allowed transitions for all considered  $12$-site systems. They all show a similar gap of $ \gtrsim 4.8 v$ and in all systems the bandwidth is approximately equal to $8$ (in the units of NN hopping, as defined in \cref{sec:LightPulseAndUnits}). Similar to the $8\!\times\!1$ case, $L^{\hat j \ket{\psi_0}}(\omega)$ comprises a small number of sharp peaks for the chains with NN-hopping only. For the chain with PBC the sharp absorption peaks are fewer and have a larger weight than for OBC as this system has more symmetries. There are more peaks for the other systems. Those peaks, when broadened, form a band. All peaks shown here and in later plots (unless mentioned otherwise) are broadened with $\sigma_\omega = 0.09$, meaning that a delta-peak in frequency is depicted as a Gaussian with standard-deviation $\sigma_\omega$.

The full time-dependent Loschmidt amplitude $L^{|\psi(t)\rangle}(\omega)$ is shown in Fig.~\ref{fig::FGRandLs} (b)-(f) for different times (color range from blue to orange), together with $L^{\hat j \ket{\psi_0}}(\omega)$ (light green). 
For the earliest time, at the onset of the pulse, only the transitions present in $L^{\hat j \ket{\psi_0}}(\omega)$ (allowed by FGR) are visible. For later times, the applied $E-$field is large enough to generate a strong out-of-equilibrium state. During the pulse, there are excitations for almost all energies in the range shown. After the pulse is over, all systems show pronounced weights at $\omega_p=11$, $2\omega_p$ and $3\omega_p$, which we can describe as  (sequential) single- double- and triple photon excitations~(similar observations for the $4\times3$ system were also made in Ref.~\cite{Maislinger2020}).

\subsection{Connection between Green's function and the Loschmidt amplitude}
\label{sec:connection_GF_and_L}

\begin{figure}[htb]
	\centering
	\includegraphics[width=\linewidth]{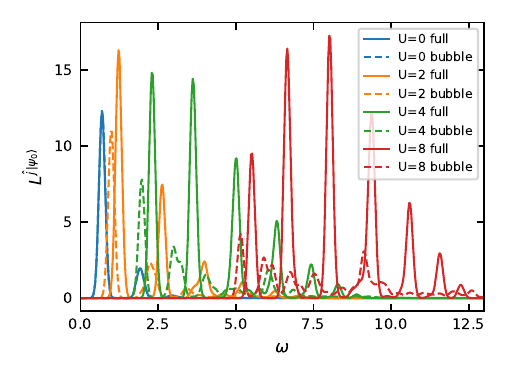}
	\caption{Optical absorption given by $L^{\hat j \ket{\psi_0}}(\omega)$ for an $8\!\times\!1$ system with NN hopping, together with the static bubble contribution (only the first term in \cref{eqn::Loschmidt_FGR_CorrFoo3} with $\vec{A}=0$) for several values of the interaction $U$. For $U=0$ the bubble contribution matches the full $L^{\hat j \ket{\psi_0}}(\omega)$ exactly (no vertex corrections).}
	\label{fig:Lj_Ns8}
\end{figure}

\begin{figure}[htb]
	\centering
	\includegraphics[width=\linewidth]{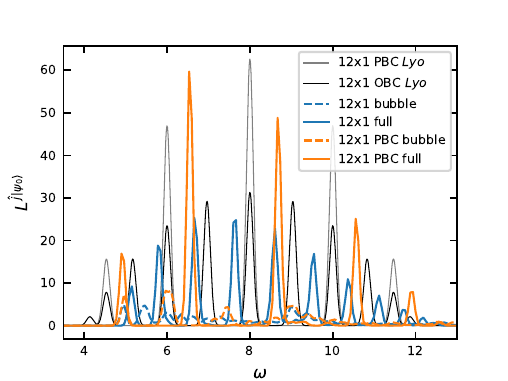}
	\includegraphics[width=\linewidth]{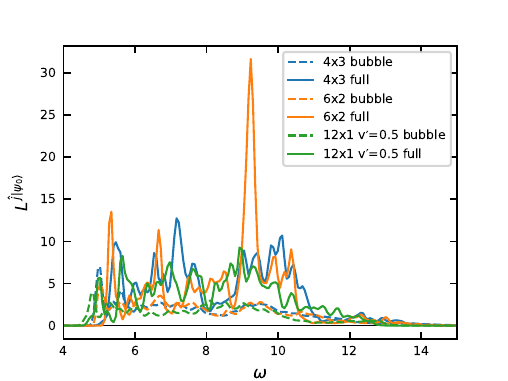}
	\caption{Optical absorption given by $L^{\hat j  \ket{\psi_0}}(\omega)$ (full lines) and its bubble contribution only (first term in \cref{eqn::Loschmidt_FGR_CorrFoo3}, dashed lines) for $N_s=12$ systems. Top panel: $12\!\times\!1$ chain with OBC and PBC together with the approximate analytical results by Lyo et. al \cite{Lyo1977} (black and gray lines). Bottom panel: $4\!\times\!3$, $6\!\times\!2$, and $12\!\times\!1$ with NNN hopping of $v'=0.5$ and OBC.}
	\label{fig:Lj_Ns12}
\end{figure}

For extended systems, exact diagonalization, which accurately captures all allowed transitions, is unfeasible.
In this case, one often uses Green's function based methods (with Feynman diagrammatics) to predict or describe optical absorption. The Loschmidt amplitude for optical absorption can also be expressed in this language.

\paragraph*{Diagrammatic expansion of $L^{\hat j \ket{\psi_0}}(\omega)$ --}  
The greater and lesser Green's functions are defined as \cite[Ch. 3]{rammer2007quantum}
\begin{equation}
\begin{array}{rcl}
G^>_{ij}(t,0) &=& -\mi      \,    \langle \hat{c}_{i}(t) \:          \hat{c}^\dagger_{j}(0) \rangle\\
G^<_{ij}(t,0) &=&  \phantom{-}\mi \, \langle \hat{c}_{j}^\dagger(0)  \: \hat{c}_{i}(t)         \rangle. 
\end{array}
\label{eqn::GgGl}
\end{equation}
The other common one-particle Green's functions (retarded-, advanced- or Keldysh-) can be constructed from the lesser and greater Green's functions by linear combinations.

We now turn to the diagrammatic expansion of the current-current correlation function $\braket{\hat j(t) j(0) }$.  
We express the current operator projected onto the $\vec A$-field through the creation and annihilation operators as $\hat j \equiv \sum_{ij} \gamma_{ij} \hat c^\dagger_i \hat c_j$.
To utilize Wick's theorem, 
the current-current correlation function
may be transformed into a contour-ordered string of operators on the Schwinger-Keldysh contour via the closed time-path contour formalism \cite{Schwinger1961, rammer2007quantum}.
This leads to 
\begin{equation}
\begin{array}{rcl}
\braket{\hat j(t) j(0) }
&=& \Tr [ \hat \rho_0\; \mathcal T_C \; \me^{-\mi \int_C \md \tau \hat H^{'}(\tau)}\; \hat j^{[+]}(0) \; \hat j^{[-]}(t)] \\
&=& \sum _{ij i'j'} \gamma_{ij} \gamma_{i'j'} \; \Tr\!\!\left[\hat \rho_0\; \mathcal T_C \;\me^{-\mi \int_C \md \tau \hat H^{'} (\tau)} \right .\\
& &  \left. \hat c^{\dagger \: [+]}_{i}(0^+) \;\: \hat c^{\phantom\dagger [+]}_j   \:   \hat c^{\dagger \: [-]}_{i'}(t+0^+) \: \hat c^{\phantom\dagger [-]}_{j'} (t)\right],
\end{array}
\label{eqn::Loschmidt_FGR_CorrFoo2}
\end{equation}
in the notation of \cite[Ch.~4.3.2]{rammer2007quantum}. That is, the superscript indices $[+]/[-]$ denote the Schwinger-Keldysh forward/backward contour and $0^+$ denotes $\lim\limits_{\epsilon\rightarrow0; \epsilon>0}\epsilon$; The perturbation $H^{'}(\tau)$ is the interacting part of the Hamiltonian together with the external perturbation in the interaction picture with respect to the kinetic term $\sum_{ij\sigma} {v_{ij}(t\!=\!0)}\,\creation{j\sigma}\annihilation{i\sigma}$  acting as the unperturbed  reference.  
The disconnected contraction of Wicks theorem vanishes as the current expectation value is zero without an external field. The other resulting terms can then be distributed into the bubble term (with a full Green's function) and vertex corrections.
\begin{equation}
\begin{array}{rcl}
\braket{\hat j(t) j(0) } &=&   \sum_{ij i'j'} \gamma_{ij} \gamma_{i'j'} \: G^>_{j'i}(t,0) \: G^<_{i'j}(0,t)  \\
& & + \: (\text{vertex corrections}).
\end{array}
\label{eqn::Loschmidt_FGR_CorrFoo3JJ}
\end{equation}

For a time-independent Hamiltonian, we can rearrange Eq.~\eqref{eqn::L_FGR1} and   express the Loschmidt amplitude for the absorption as 
\begin{equation}
L^{\hat j \ket{\psi_0}}(t) = \bra{\psi_0} \hat j(t) j(0) \ket{\psi_0},
\label{eqn::Loschmidt_FGR_CorrFoo1}
\end{equation} 
where we inserted $\mathbf{1} = \me^{-\mi t \hat H } \me^{\mi t \hat H }$ with $\hat{H}=\hat{H}(t=0)$ before the first current operator in ~\eqref{eqn::L_FGR1}, replaced $\tau$ with $t$,  and  set the ground-state energy to zero. 
Notice that for an arbitrary state (or more general a mixture of states) modifications to the above expression would be necessary. 



The leading order terms (in $\vec{A}$) for a diagrammatic expansion of the Loschmidt amplitude for optical absorption  are thus given by 
\begin{equation}
\begin{array}{rcl}
L^{\hat j \ket{\psi_0}}(t) 
&=&   \sum_{ij i'j'} \gamma_{ij} \gamma_{i'j'} \: G^>_{j'i}(t) \: G^<_{i'j}(-t)  \\
& & + \: (\text{vertex corrections}) \:+ \:  {\mathcal{O}}(\vec{A})
%
\end{array}
\label{eqn::Loschmidt_FGR_CorrFoo3}
\end{equation}
where $G^<_{ij}(t)/G^>_{ij}(t)$ is the lesser/greater equilibrium Green's function and the vertex corrections are understood with respect to $U$, not $\vec{A}$.
The above expression links the FGR-given absorption at short times with the quasiparticle picture of Refs.~\cite{Kauch2020Disorder,Maislinger2020}, where the description using the one-particle Green's function (spectral function) was used to discuss the presence or absence of impact ionization. In this Green's function based  quasiparticle picture light can be absorbed only if the pulse frequency is larger than the gap and smaller than the total bandwidth of the spectral function. This approach corresponds to taking only the bubble contribution into account. In the following we analyze for which systems such an approach fails.

In Fig.~\ref{fig:Lj_Ns8} we show the full Loschmidt amplitude for absorption $L^{\hat j \ket{\psi_0}}$ together with the bubble contribution (first term in Eq.~\eqref{eqn::Loschmidt_FGR_CorrFoo3}) for the $8\!\times\!1$ systems with different interaction values. In the case of $U=0$ there are no vertex corrections and the bubble contribution is equal to the full function. In all other cases, the vertex corrections change the result quite significantly. What is described reasonably well already by the bubble term (and thus by the quasi-particle picture) are the size of the gap and the bandwidth. The vertex corrections lead to a slight narrowing of the bandwidth and also to a slightly larger gap. However, the sharp, almost equally spaced peaks in the full $L^{\hat j \ket{\psi_0}}(\omega)$ are described only by the inclusion of vertex corrections. The bubble term completely fails to reproduce the correct position of the peaks as well as their weight.

The almost equal-distance distribution of the absorption peaks can be understood by looking at the one-dimensional systems at strong coupling. As discussed e.g. by  Lyo {\it et al.}~\cite{Lyo1977}, the major contribution to the photoabsorption is given by the following process: At half filling in the large $U$ limit each site is occupied by exactly one electron. Incoming light can create through the current operator a hole and a double occupancy at a neighboring site. The hole and the double occupancy (doublon) can subsequently propagate through the system. The doublon-hole pair has a total momentum of zero. The doublon and the hole each carry a momentum of $k$, which takes only several discrete values for small chains. These values are reflected in the different peaks visible in the absorption spectrum. Lyo et al. \cite{Lyo1977} showed that for the large $U$ limit in the antiferromagnetic phase the optical conductivity (which is directly related to the Loschmidt amplitude; see next paragraph) has in one dimension the following form
\begin{equation}
\sigma^R(\omega) = \frac{4 \pi v^2}{\omega} \sum_{k} \sin^2(k) \; \delta\left(U -4v\cos(k) -  \omega\right)
\label{eqn:Lyo}
\end{equation}
where $k$ is the momentum of the doublon-hole pair and $v$ is the NN hopping (set to $1$ in this paper).  For PBC the allowed momenta are $k=\frac{2 \pi n}{N_s}$ where $n\in [0, N_s-1]$. For OBC we have instead $k=\frac{ \pi n}{N_s}$ which corresponds to more different energies. This explains why there are twice as many peaks for OBC as for PBC in~\cref{fig::FGRandLs}(a). 

In the top panel of \cref{fig:Lj_Ns12} we show   $L^{\hat j \ket{\psi_0}}(\omega)$ (bubble and full), together with the from \cref{eqn:Lyo} predicted peak positions for the $N_s=12$ Hubbard chain, for $U=8$ for both OBC and PBC. Again we see that although the gap and bandwidth are well reproduced by the bubble contribution (dashed lines in \cref{fig:Lj_Ns12}), the sharp peak structure is completely determined by the vertex corrections (blue and orange full lines in \cref{fig:Lj_Ns12}). The optical spectral weight is also considerably enhanced by vertex corrections. The analytical results of Lyo et al. do not, in this case, give the correct gap size. They would predict a gap of $U-4$ without finite-size effects. For a $12$-site system with PBC the gap is predicted to be $U-3.46$ since for $k=0$ the matrix element (sine function in \cref{eqn:Lyo}) is zero. For our PBC results the lowest $\omega$ for which the system can absorb energy is larger: $\omega=4.9$ (for OBC it is $5.15$,  also larger than the predicted value of $U-3.86$). 
The difference is not solely due to the not fully applicable large $U$ expansion as also for larger $U$ values we do not get the predicted gap (see~\cref{app:Lyo} where the results for $U=8 \cdot 2^n$ for $n \in [0,1,2,3]$ are shown in~\cref{fig:Scaling_12x1pBC} and a brief explanation is given). The overall structure of the spectrum is, however, well reflected by the approximate expression in \cref{eqn:Lyo}. 

In the bottom panel of \cref{fig:Lj_Ns12} we show the bubble contribution (dashed lines)  and full  $L^{\hat j \ket{\psi_0}}(\omega)$ (full lines) for $12$-site systems with higher connectivity. Contrary to the case of NN-hopping chains shown in the top panel, the full spectrum does not consist of few well-separated peaks. However, also here the vertex corrections contain a large weight that is differently distributed than in the bubble and there are additional peaks stemming solely from vertex corrections. The gap and bandwidth are already well predicted by the bubble part, although the gap is slightly larger when vertex corrections are included.


\subsection{Connection to optical conductivity}
There is an intimate relationship between the retarded current-current correlation function $K^R(t,t')$ and the Loschmidt amplitude $L^{\hat j \ket{\psi}_0}$.
The former describes how the current responds to the perturbation of the system with a classical $E-$field within linear response theory. The latter tells us at which energies the system can absorb energy in lowest-order perturbation theory in $E$. They are related as
\begin{equation}
\begin{array}{rcl}
K^R(t,0) &=& \mi \theta(t) \, \braket{[\hat j(t), \hat j(0)]}_0 \\
&=& - 2 \theta(t) \, \Im L^{\hat j \ket{\psi}_0}(t).
\end{array}
\label{eqn::OptCondTime}
\end{equation} 
Note that for an arbitrary initial state (not the ground state used here) this relation will have to be modified. We show in the \cref{App:Properties} that in frequencies the (antisymmetric) imaginary part of the retarded current-current correlation function is for positive frequencies exactly twice the Loschmidt amplitude for the state $\hat j \ket{\psi}_0$:  
\begin{equation}
L^{\hat j \ket{\psi}_0}(\omega > 0) = 2 \: \Im K^{R}(\omega > 0 ).
\label{eqn::OptCondFreq}
\end{equation}


\section{Results for the generalized Loschmidt amplitude}
\label{sec:GL_all}

\subsection{Generalized Loschmidt amplitude for the double occupancy and energy}
\label{sec:GL_docc}

\begin{figure*}
	\centering
	\subfloat[$4\!\times\!3$ total]{
		\includegraphics[width=\TMPWW]{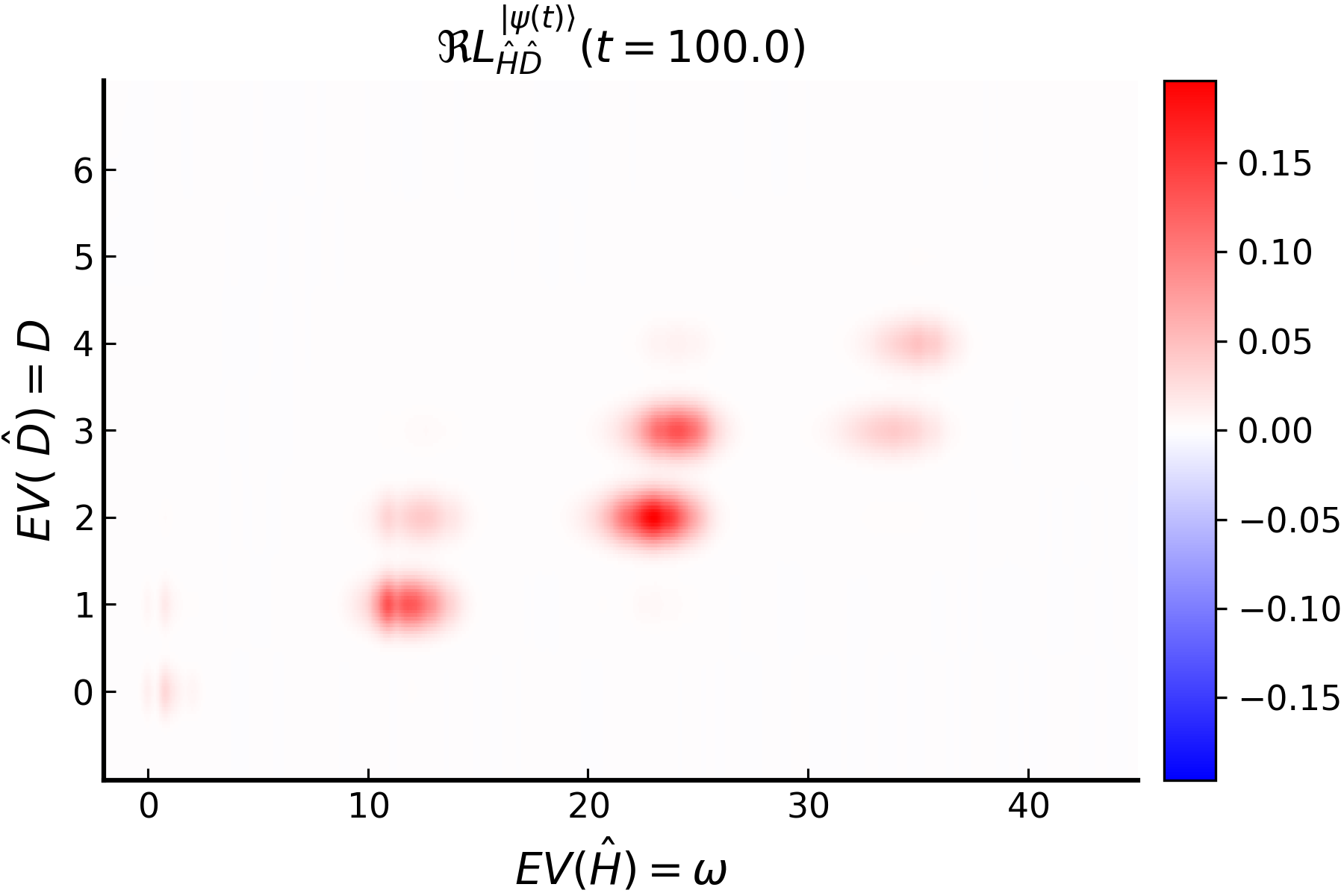}
	}
	\subfloat[$4\!\times\!3$]{
		\includegraphics[width=\TMPWW]{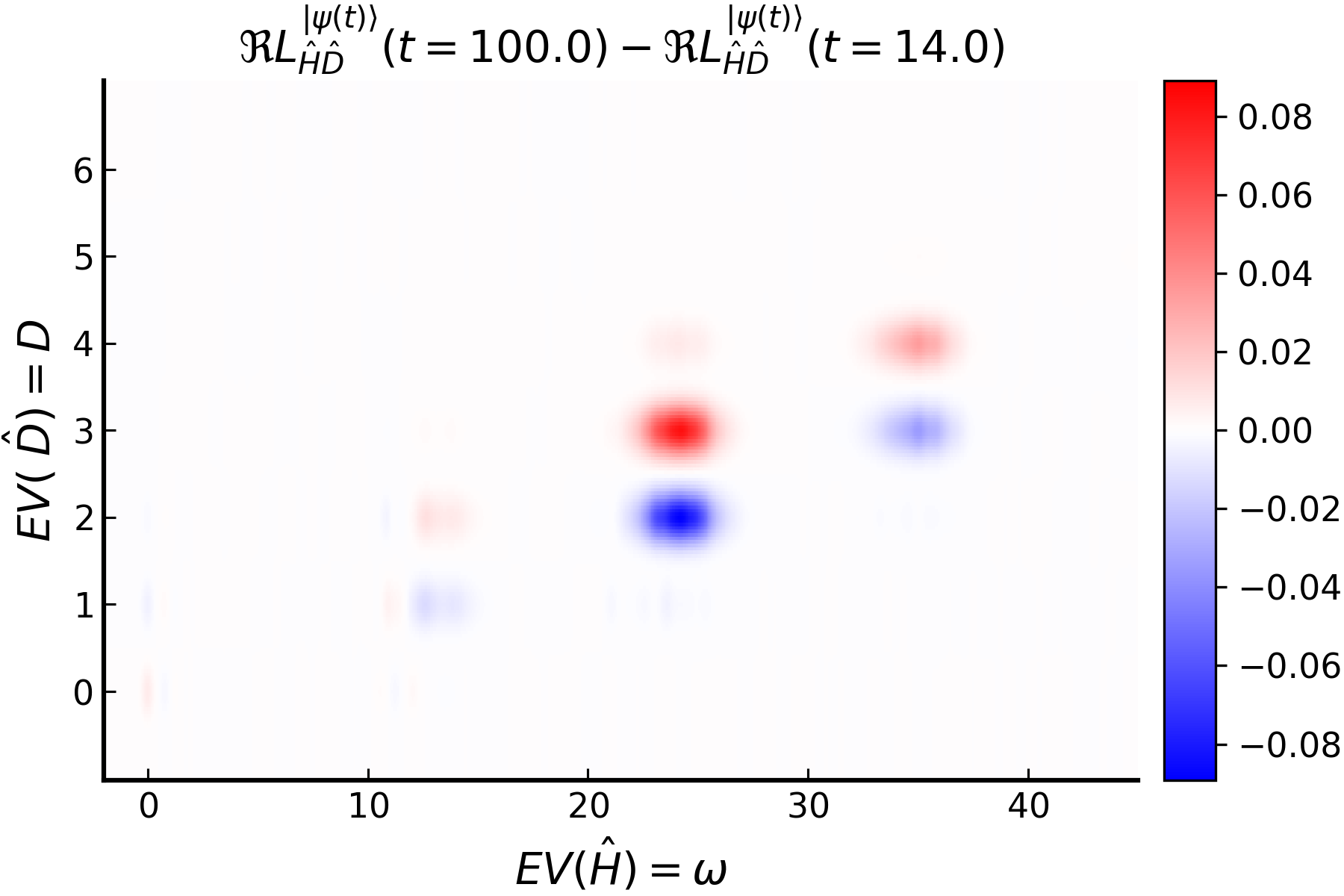}
	}
	\hspace{0mm}
	\subfloat[$6\!\times\!2$]{
		\includegraphics[width=\TMPWW]{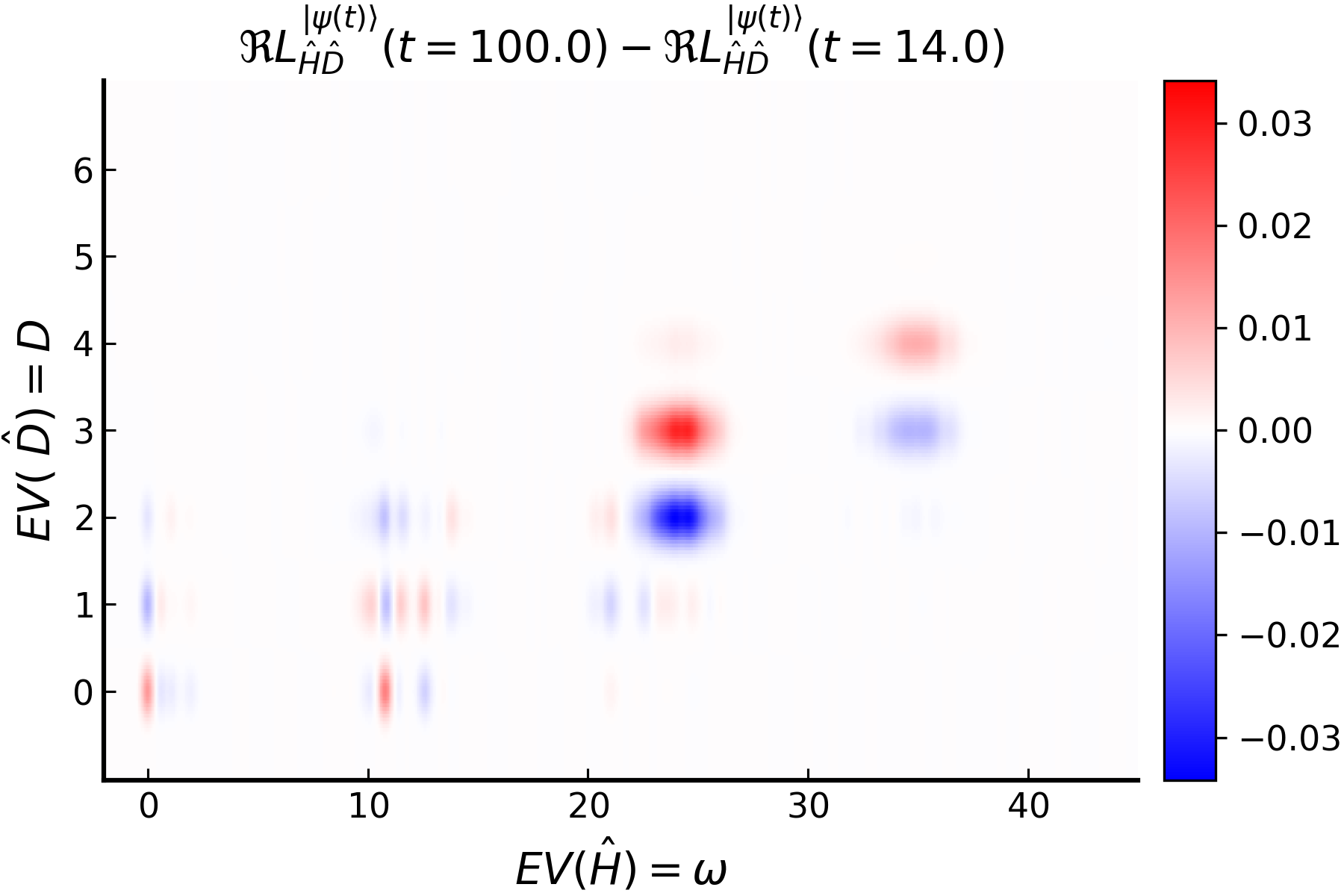}
	}
	\subfloat[$12\!\times\!1$ OBC]{
		\includegraphics[width=\TMPWW]{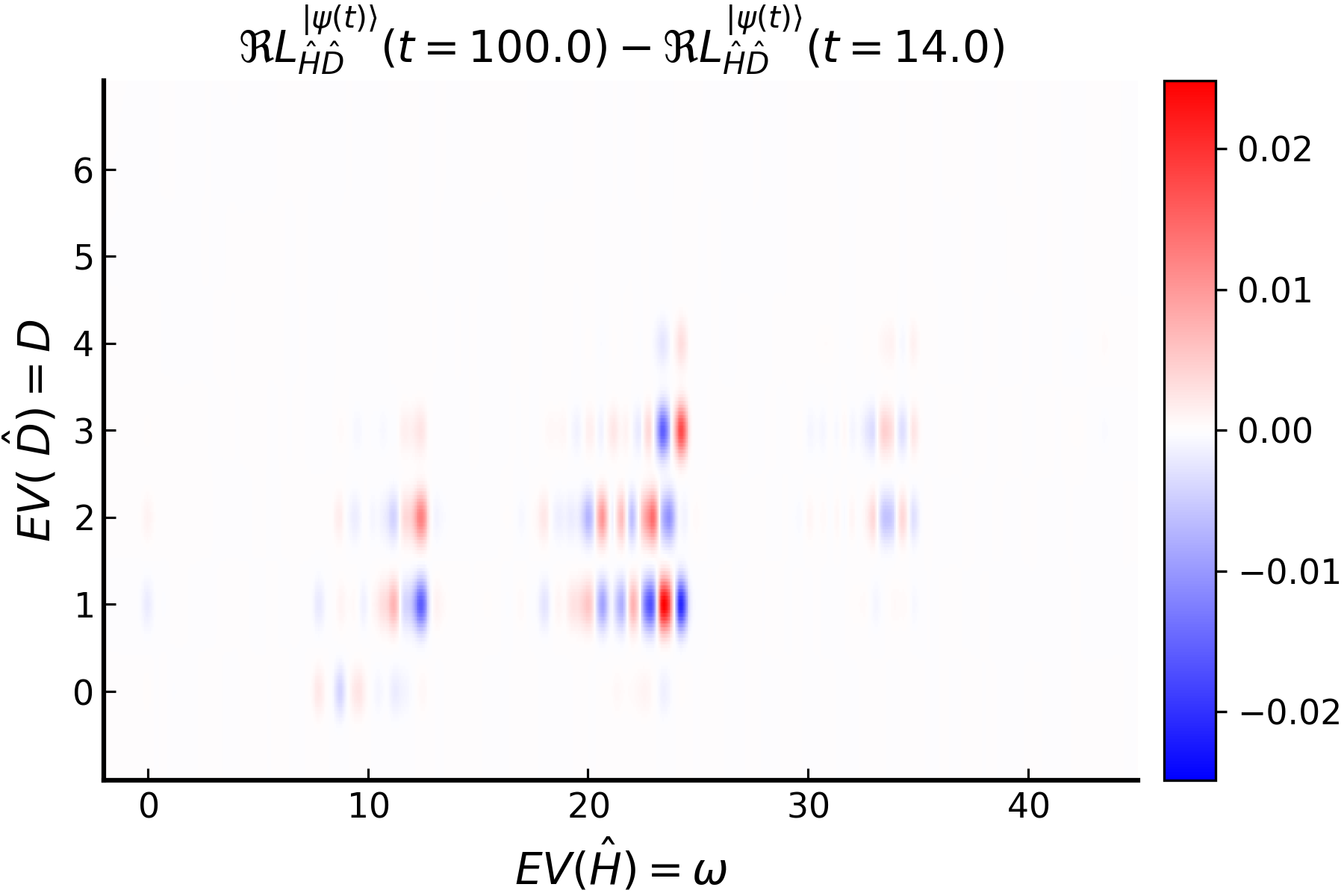}
	}
	\hspace{0mm}
	\subfloat[$12\!\times\!1$ PBC]{
		\includegraphics[width=\TMPWW]{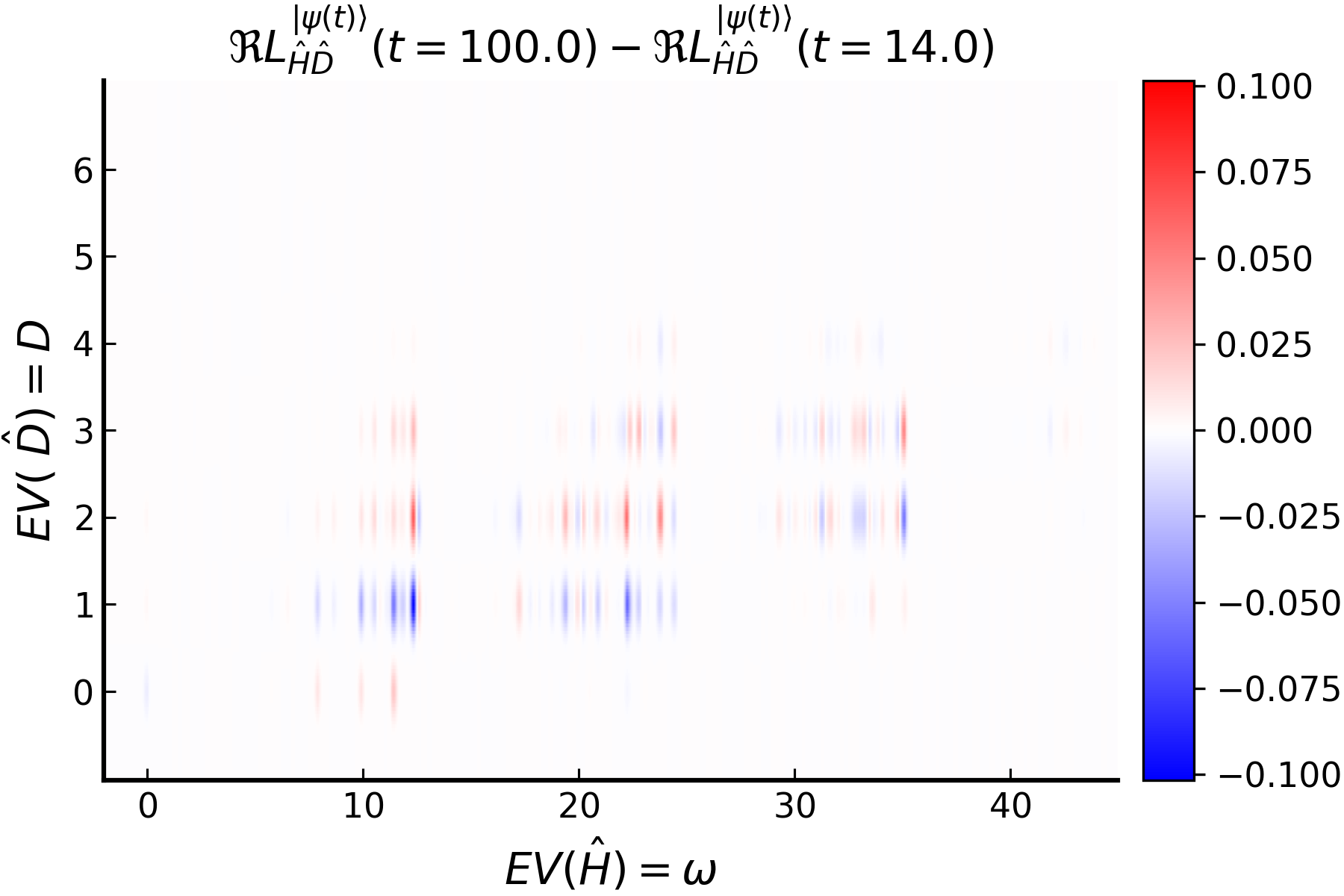}
	}
	\subfloat[$12\!\times\!1$ $v^\prime=0.5$]{
		\includegraphics[width=\TMPWW]{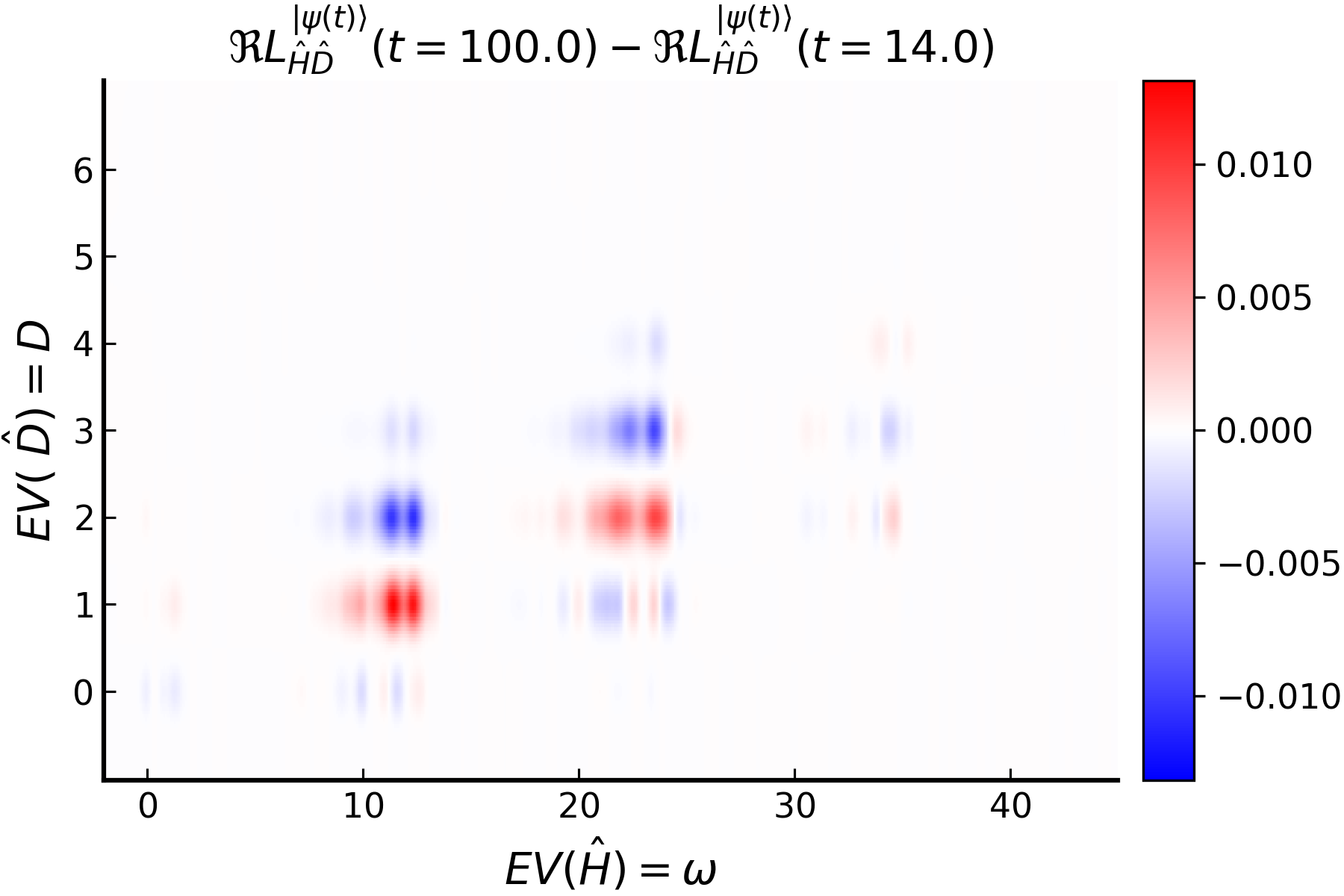}  
	}
	\caption{Generalized Loschmidt amplitude for the $N_s=12$ site systems, allowing us to better analyze the dynamics of the double occupations. In (a) the real part of $L^{\ket{\psi(t)}}_{\hat H \hat D}$ is shown for $t=100$. Plots (b)-(f) show the difference between $L^{\ket{\psi(t)}}_{\hat H \hat D}$ long after the pulse ($t=100$) and shortly after the pulse ($t=14$). For all systems shown a consistent energy broadening of $\sigma_\omega=0.18$ 
		and double occupancy eigenvalue broadening of $\sigma_d=0.1$ has been employed.}
	\label{fig::GL_Docc}
\end{figure*}

In the context of impact ionization it is interesting to know which energy-states are responsible for the long-time dynamics of the double occupancy after the $E-$field pulse is over. To this end we consider the generalized Loschmidt amplitude as defined in Eq.~\eqref{eqn::LAB_def2} with  $\hat A = \hat H(0)$ and $\hat B = \hat D \equiv \sum_i {\hat n_{\uparrow i} n_{\downarrow i} }$. The double-occupancy operator $\hat D$ has the eigenvalues $D\equiv EV(\hat{D}) = \{0, 1, ... , N_s/2\}$ for half-filling. This leads to 
\begin{equation}
\begin{array}{rcl}
L^{\ket{\psi(t)}}_{\hat H \hat D}(\omega, D) &=&  (2 \pi)^2 \sum_n \sum_m 
\braket{\psi(t) | E_n} \braket{ E_n | D_m}   \\ 
&&\times\braket{D_m | \psi(t)}  \delta(\omega - E_n) \, \delta(D -D_m).
\end{array}
\label{eqn::LHD_def2}
\end{equation} 
During the pulse, the Hamiltonian is time-dependent and spectral weight can be shifted between different eigenvalues of $\hat H$ (named $\omega\equiv EV(\hat H)$). In  \cref{fig::GL_Docc}(a) we show the real part of the generalized Loschmidt amplitude $L^{\ket{\psi(t)}}_{\hat H \hat D}(\omega, D)$ for a time $t=100$ long after the pulse. We find an almost linear relationship between the double-occupancy eigenvalues and the eigenenergies with a slope that is close to the strong-coupling limit of $1/U$. 

After the pulse (centered at $t_p=8$), the Hamiltonian is time-independent and the $L^{|\psi(t)\rangle}(\omega)$ also becomes static. The generalized Loschmidt amplitude, on the other hand, stays a dynamic quantity also after the pulse is over. However, given that
\begin{equation}
\int \md D \: L^{\ket{\psi(t)}}_{\hat H \hat D}(\omega, D) = 2 \pi L^{|\psi(t)\rangle}(\omega)
\end{equation}
there can only be dynamics with respect to $D$. In other words, there can be a redistribution of spectral weight along the $EV(\hat{D})-$axis but not along the $EV(\hat{H})-$axis after the pulse. Hence, if after the pulse there is an overall trend to a redistribution of spectral weight towards larger double occupancy values, we witness impact ionization. From our generalization of the Loschmidt amplitude, one can tell which energy states are responsible for the double-occupancy dynamics. This reflects the fact that the generalized Loschmidt amplitude is a spectral decomposition with respect to two operators at the same time. 

\subsubsection{Comparison of  $12$-site systems with different geometry}

We find for all the systems under consideration that the single-photon excitations  ($EV(\hat H)\approx \omega_p$) predominately consist of $EV(\hat D) = 1$ and $2$ and the double-photon excitations of $EV(\hat D) = 2$ and $3$. The $4\!\times\!3$ and $6\!\times\!2$ systems show impact ionization (also seen in Ref.~\onlinecite{Kauch2020Disorder} in the rise of the double occupancy as a function of time for times after the pulse is over). This finding is well captured by the generalized Loschmidt amplitude in \cref{fig::GL_Docc}{(b)-(c)}, where we show a difference between the $\Re L_{\hat H \hat D}$ shortly after the pulse (at  $t=t_p+3\sigma_p=14$) and at a  much later time ($t=100$). We see that for the same energy eigenstates, the number of double occupancies increases. The strongest change in the double occupation comes from the energy states at double-photon excitations ($EV(\hat H)\approx 22$) or even triple-photon excitations ($EV(\hat H)\approx 33$) and not from single-photon excitations (also seen in Ref.~\onlinecite{Maislinger2020}). This behavior depends both on the concrete system under consideration and on the pulse frequency. In the $4\times2$ Hubbard cluster, single-photon excitations also significantly contribute to the impact ionization for optimal parameters (see~\cref{fig:GL4x2vd07} in \cref{App:Scan}). In the $4\times3$ cluster, single-photon excitations are only important for larger pulse frequencies, where we observe weaker impact ionization (see next section).

\begin{figure*}
	\centering
	\subfloat[$4\times3$, $\omega_p=10$]{
		\includegraphics[width=\TMPWW]{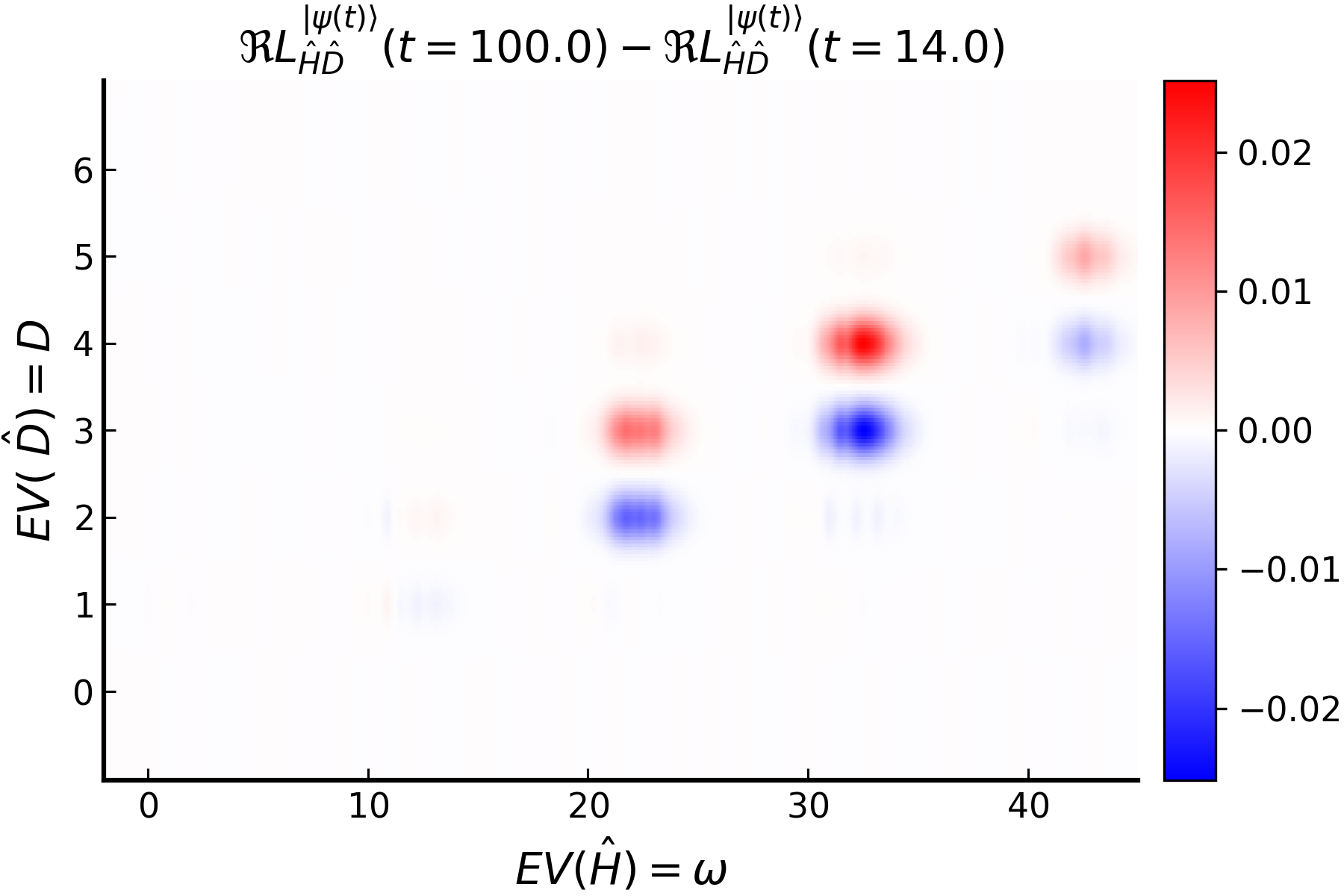}
	}
	\subfloat[$4\!\times\!3$, $\omega_p=10$, integrated]{
		\includegraphics[width=\TMPWW]{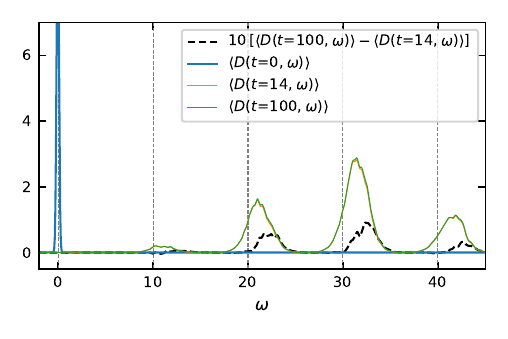}
	}
	\hspace{0mm}
	\subfloat[$4\times3$, $\omega_p=12$]{
		\includegraphics[width=\TMPWW]{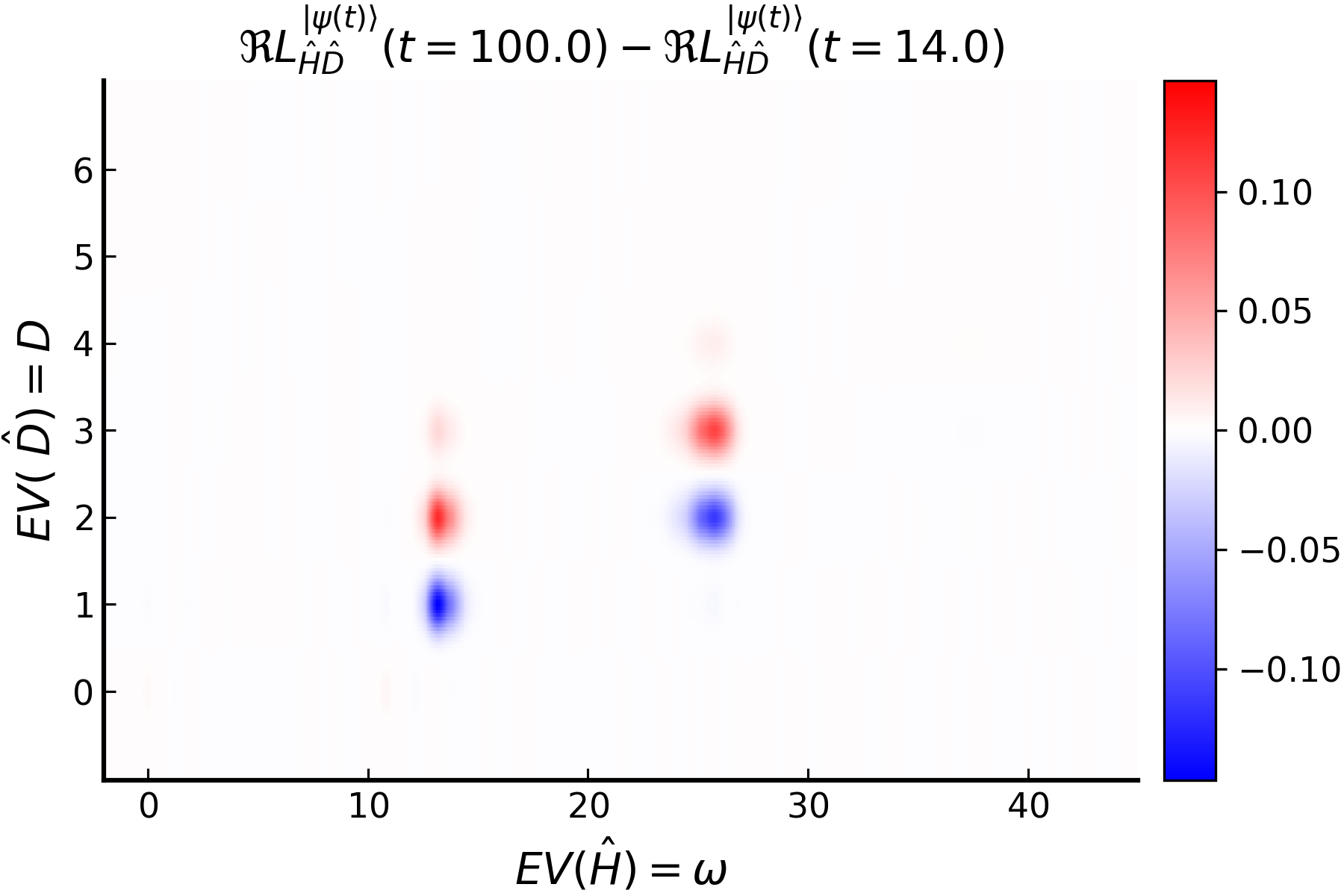}
	}
	\subfloat[$4\!\times\!3$, $\omega_p=12$, integrated]{
		\includegraphics[width=\TMPWW]{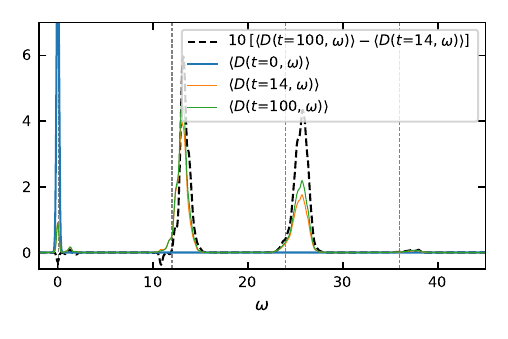}
	}
	\hspace{0mm}
	\subfloat[$4\times3$, $\omega_p=13$]{
		\includegraphics[width=\TMPWW]{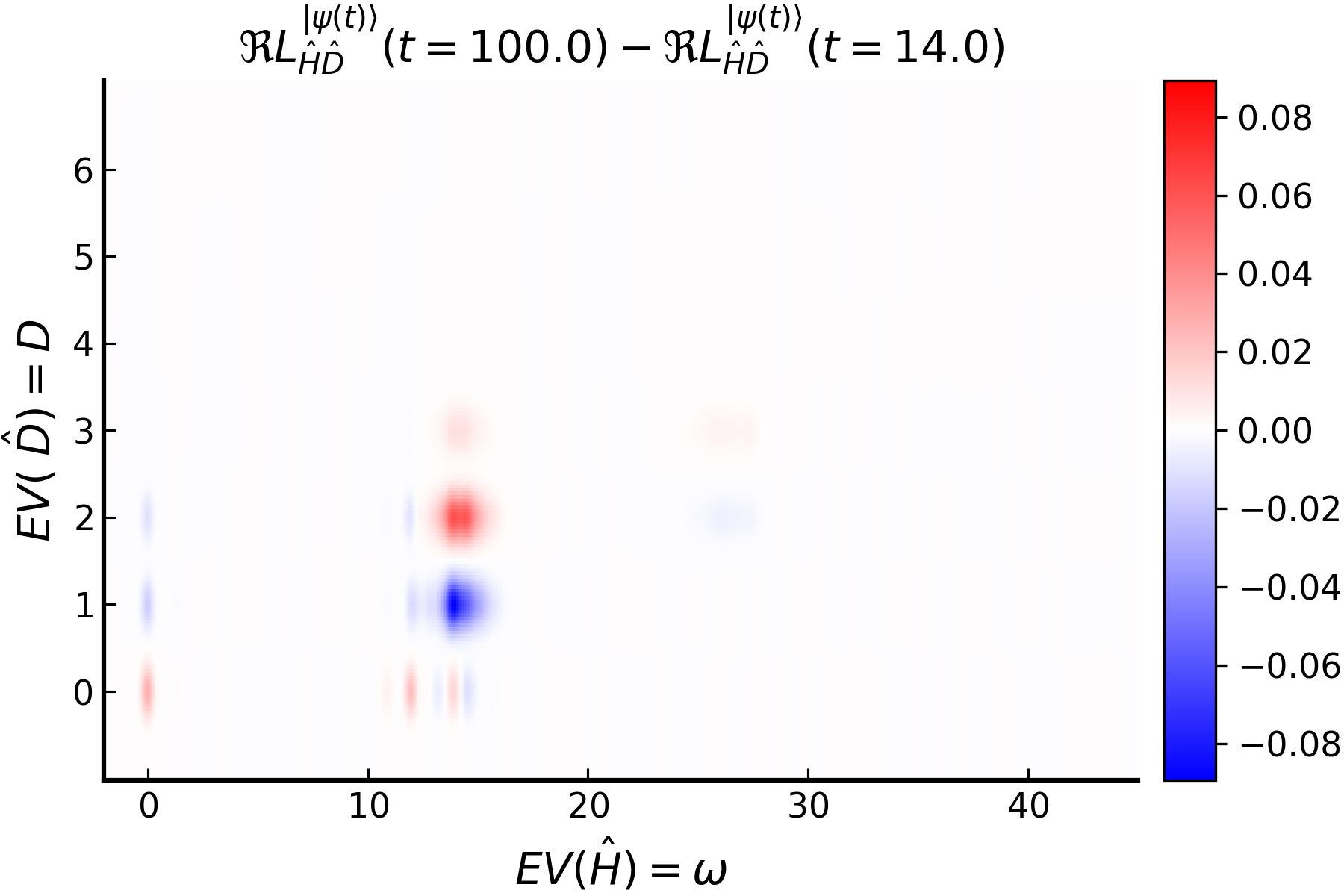}
	}
	\subfloat[$4\!\times\!3$, $\omega_p=13$, integrated]{
		\includegraphics[width=\TMPWW]{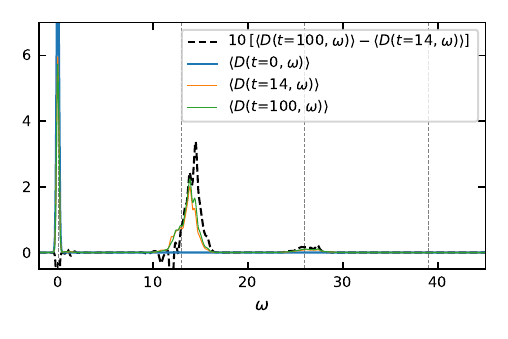} 
	}
	\caption{Generalized Loschmidt amplitude for the $4\times 3$ system. Left column: the difference between $L^{\ket{\psi(t)}}_{\hat H \hat D}$ long after the pulse ($t=100$) and shortly after the pulse ($t=14$). Right column: average double occupation per energy eigenvalue $\braket{D(t,\omega)}$, as defined in \cref{eqn::av_docc}, at different times. Black dashed curves show the difference of $\braket{D(t,\omega)}$ between $t=100$ and $t=14$ (multiplied by $10$ to improve visibility). Dotted vertical lines indicate integer multiples of the pulse frequency $\omega_p$. Different rows show results for different pulse frequencies: (a,b) $\omega_p=10$, (c,d) $\omega_p=12$, (e,f) $\omega_p=13$. For all systems shown a consistent energy broadening of $\sigma_\omega=0.18$ 
		and double occupancy eigenvalue broadening of $\sigma_d=0.1$ has been employed.}
	\label{fig::GL_Docc_freqs}
\end{figure*}

An inverse effect to impact ionization (impact deionization) can be seen in the double occupation dynamics of the frustrated $12\!\times\!1$ system shown in \cref{fig::GL_Docc}{(f)}. Here both the single and the double-photon excitations are important to the dynamics. However, the generalized Loschmidt amplitude reveals a decrease of double occupations.

For the $12$-site chains with only NN hopping and OBC or PBC (\cref{fig::GL_Docc} {(d)-(e)} there is no net increase in the double occupancy between $t=14$ and $t=100$. There is, however, still strong dynamics of the double occupancy visible. For example for $12\!\times\!1$ in \cref{fig::GL_Docc}(d) we see at $EV(\hat H) \approx 23.5$  a reduction of double occupancy whereas at  $EV(\hat H) \approx 22.8$ there is an increase. Summing up all contributions gives a cancellation and there is no overall increase in double occupation (which can be interpreted as effectively no impact ionization). This was also found to be the case in~Ref.~\cite{Kauch2020Disorder}, where the time dependence of double occupancy after the pulse was shown for precisely the same systems and parameters as in our work (for a parameter scan for $12$-site systems, see Appendix~\ref{App:Scan}).  

\subsubsection{$4\times3$ system for different pulse frequencies}


In the following we will focus on the system showing the strongest impact ionization, namely the $4\times3$ system. The results in Fig.~\ref{fig::GL_Docc} were obtained for pulse frequency $\omega_p=11$ which is very close to optimal for observing impact ionization (see  Fig.~\ref{fig:parameter_scan} in Appendix~\ref{App:Scan} and also Ref.~\cite{Maislinger2020}). In Fig.~\ref{fig::GL_Docc_freqs} we present the generalized Loschmidt amplitude for several other pulse frequencies $\omega_p$ in the range where according to our parameter scan (cf. Fig.~\ref{fig:parameter_scan}) impact ionization occurs, i.e., $\omega_p\in[10,13]$. In the left column of Fig.~\ref{fig::GL_Docc_freqs} [in the plots a), c), and e)], the difference between $\Re L_{\hat H \hat D}$ long after the pulse ($t=100$) and shortly after the pulse ($t=14$) for three different pulse frequencies is shown, analogously to Fig.~\ref{fig::GL_Docc}. To highlight the changes in the double occupation per energy eigenvalue, we show, in the right column of  Fig.~\ref{fig::GL_Docc_freqs} [in the plots (b), (d), and (f)], the following integral  
\begin{equation}
\braket{D(t,\omega)} = \frac{1}{2\pi} \int \md D \: D \: L^{\ket{\psi(t)}}_{\hat H \hat D}(\omega, D).
\label{eqn::av_docc}
\end{equation}
This quantity represents an energy-resolved double occupancy. It is related to the expectation value of the double occupancy as:
\begin{equation}
	\braket{\hat D(t)} = \frac{1}{2\pi} \int \md \omega \, \braket{D (t,\omega)}.
\end{equation}
In Fig.~\ref{fig::GL_Docc_freqs} it is shown for three times ($t=0$, $t=14$, and $t=100$). Also, the difference between values at long times, $t=100$, and shortly after the pulse, $t=14$, is shown (multiplied by a factor of $10$ to be more visible in the plot). We see that, depending on pulse frequency, the biggest changes in the double occupation happen at different energies. For $\omega_p=10$, the strongest contribution to impact ionization comes from $EV(\hat{H})\approx20$ and $30$, which corresponds to double- and triple-photon excitations, whereas for $\omega_p=12$ and particularly for $\omega_p=13$, the single-photon excitations contribute the most to the increase of the double occupation.  

In order to qualitatively understand this behaviour, let us refer to the quasi-particle description of impact ionization. In this picture, impact ionization can happen if the energy of the photon is bigger than twice the gap (the excess kinetic energy of the fist electron-hole pair is used to create a second electron-hole pair). A nice pictorial view of the possible processes is shown in Fig. 3 of Ref.~\cite{Maislinger2020}. The optical gap of the $4\times3$ system is, as can be seen in \cref{fig:Lj_Ns12}, slightly bigger than the one-particle gap in the spectral functions presented in \cref{fig:EqGF}, and is $\approx 5$. In agreement with that, we find impact ionization to occur for pulse frequencies $\gtrapprox 10$ (see \cref{App:Scan}). Intuitively, the single-photon processes should dominate in the entire range of the pulse frequencies for which we see impact ionization. For the almost optimal frequency for impact ionization, $\omega_p=11$, as well as for $\omega_p=10$, this is not the case. The time dependence of the Loschmidt amplitude in \cref{fig::FGRandLs} shows that for $\omega_p=11$ already during the pulse, at $t=8$, the system is dominated by excitations at energies corresponding to double-photon excitations, with significant contributions from triple-photon exctations, which were sequentially generated from the single-photon excited states allowed by FGR at the beginning of the pulse. The subsequent time evolution of double occupation hence also happens in the double- and triple-photon energy range. For  $\omega_p=13$ the $4\times 3$ system absorbs much less energy (the FGR allowed states have much less weight at $\omega=13$, cf. \cref{fig::FGRandLs}) and therefore the sequential absorption in the double-photon energy range does not take place and the subsequent dynamics of double occupation happens in the single-photon energy range. 

The dominance of the energy range corresponding to double-photon excitations in the double occupation dynamics is surprising and likely  specific for these particular $12$-site systems with impact ionization. For the smaller $4\times2$ systems, single-photon processes are important in the entire range of pulse frequencies for which impact ionization is present (see also \cref{App:OtherFigures4x2}). 

\subsection{Spin excitations}
\label{sec:SpinExcitations}

%
%
%
%
%

Before we apply the generalized Loschmidt amplitude to spin excitations, let us first investigate the time dependence of the spin-spin correlation function and spin-energy for $12$-site systems after the light pulse. The main motivation to do so lies in our earlier work, Ref.~\cite{Kauch2020Disorder}, where it was shown that disorder and next-nearest-neighbor hopping enhance impact ionization in small Hubbard clusters. Moreover, one-dimensional chains do not show any significant impact ionization. This suggested that if a system has the tendency to order magnetically, excess kinetic energy may first break up this order or fluctuations, i.e., excess energy is transferred to magnons or paramagnons. It was conjectured that this could be detrimental to impact ionization. 
To verify this proposition we investigate the spin-spin correlation function 
\begin{equation}
\begin{array}{rcl}
C_{ij}(t,t') &=& \braket{\hat S_z^{i}(t) \hat S_z^{j}(t') },
\end{array}
\label{eqn::SpinCorrelationFunction}
\end{equation}
where in our units $\hat{S}_z^i = \frac{1}{2} (\hat n_{\uparrow}^i  - \hat n_{\downarrow}^i)$; as well as the corresponding Kubo susceptibility (shown in Appendix~\ref{App:Spin})
\begin{equation}
\begin{array}{rcl}
\chi^R_{ij}(t,t') &=&- \mi \theta(t-t') \braket{\left[\hat S_z^{i}(t), \hat S_z^{j}(t')\right] } \\
&=& \:\,2 \, \theta(t-t') \, \Im C_{ij}(t,t').
\end{array}
\label{eqn::SpinSusc}
\end{equation}
Furthermore, as a measure of the tendency for spin-order we also consider the expectation value of the Heisenberg Hamiltonian
\begin{equation}
E_{\mathrm{H}}(t) = \bra{\psi(t)} \hat{H}_{\mathrm{H}} \ket{\psi(t)}  
\label{eqn::EnergyHeisenberg2}
\end{equation}
with 
\begin{equation}
\begin{array}{rcl}
\hat H_{\mathrm{H}} &=& \sum_{i > j} J_{ij} \left(\hat{S_x}^{i} \hat{S_x}^{j} + \hat{S_y}^{i} \hat{S_y}^{j} + \hat{S_z}^{i} \hat{S_z}^{j}\right).
\end{array}
\label{eqn::EnergyHeisenberg}
\end{equation}
This is motivated by the fact that in the limit of large interaction values $U$, the (static) Hubbard model can be mapped onto the Heisenberg model by the Schrieffer-Wolff transformation  (see, e.g., \cite[Chapter 2.2]{altland2006condensed}). The spins at site $i$ and $j$ couple due to super-exchange with a coupling constant given by $J_{ij} =  4 \frac{v_{ij}^2}{U}$. In \cref{eqn::EnergyHeisenberg} all three terms give the same contribution due to $SU(2)$ symmetry. We utilized this fact to speed up the calculations. 
Although the Schieffer-Wolff mapping is far from being exact for $U=8$ used in this paper, it nonetheless gives an intuitive understanding of the properties of the Hubbard clusters.
For example, for $J>0$ (for NN-hopping), neighboring spins can lower $E_{\mathrm{H}}$ by aligning antiparallelly.

In the following we present the results for $12$-site clusters with open boundary conditions (OBC) unless explicitly stated otherwise.


\begin{figure*}
	\centering
	\subfloat[$4\!\times\!3$]{
		\begin{picture}(250,170)
		\put(0,0){\includegraphics[width=\TMPWW, trim=0cm 4mm 0cm 0.0cm, clip=true]{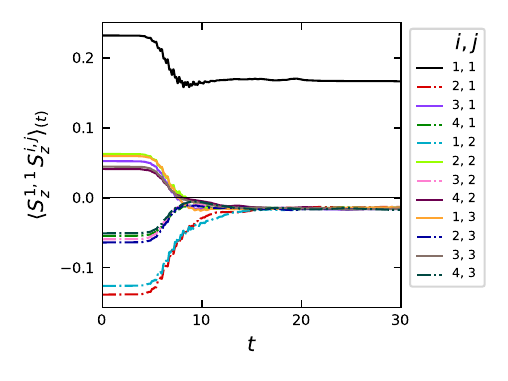}}
		\put(110,80){\includegraphics[width=2.5cm, trim=2.2cm 1.5cm 1.8cm 1.33cm, clip=true]{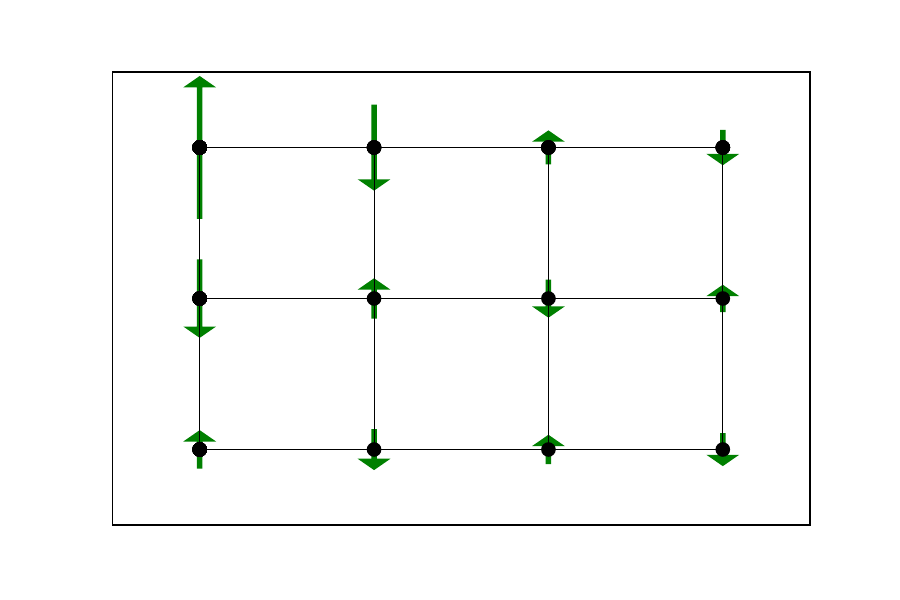}}
		\end{picture}
	}
	\hspace{0mm}
	\subfloat[$6\!\times\!2$]{
		\begin{picture}(250,170)
		\put(0,0){\includegraphics[width=\TMPWW, trim=0cm 4mm 0cm 0.0cm, clip=true]{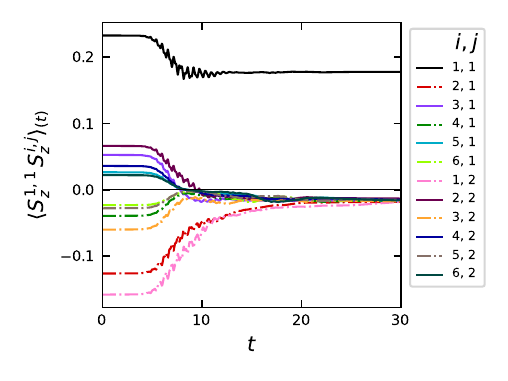}}
		\put(110,80){\includegraphics[width=2.5cm, trim=2.2cm 1.5cm 1.8cm 1.35cm, clip=true]{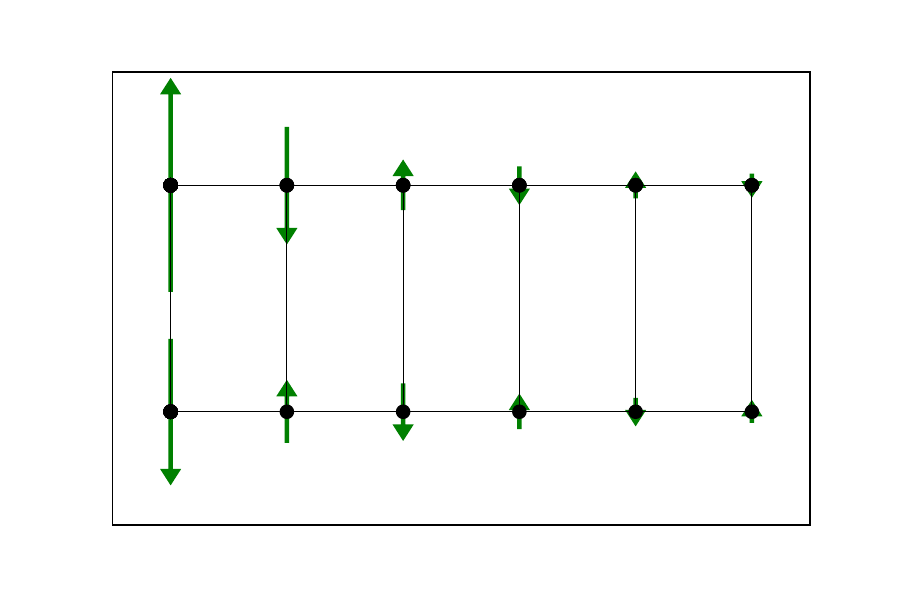}}
		\end{picture}
	}
	\subfloat[$12\!\times\!1$]{
		\begin{picture}(250,170)
		\put(0,0){\includegraphics[width=\TMPWW, trim=0cm 4mm 0cm 0.0cm, clip=true]{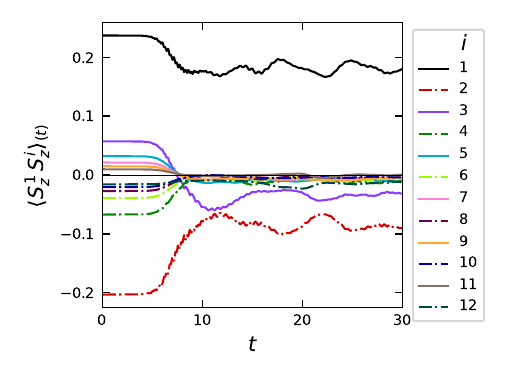}}
		\put(110,92){\includegraphics[height=1.3cm, trim=2.2cm 1.5cm 1.8cm 1.35cm, clip=true]{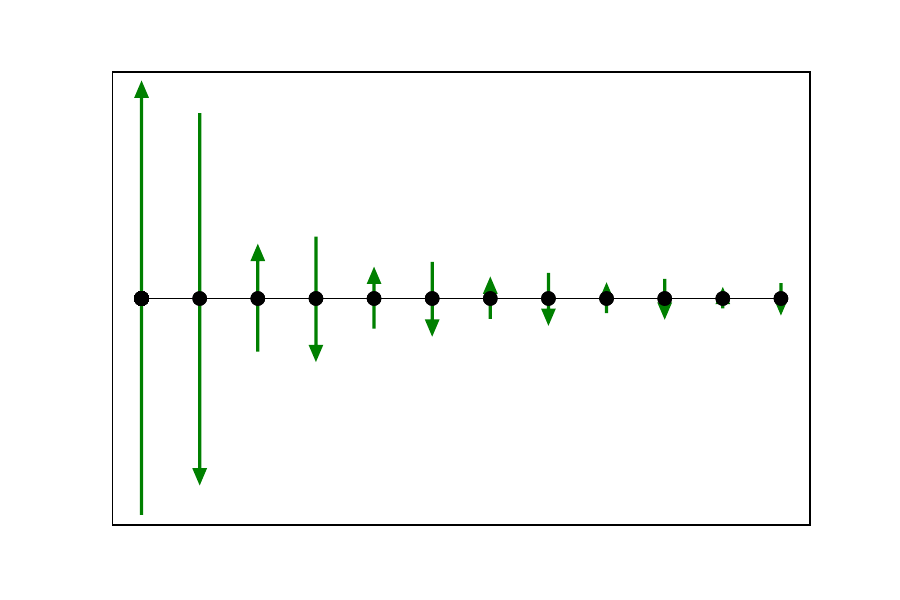}}
		\end{picture}
	}
	\hspace{0mm}
	\subfloat[$12\!\times\!1$ PBC]{
		\begin{picture}(250,170)
		\put(0,0){\includegraphics[width=\TMPWW, trim=0cm 4mm 0cm 0.0cm, clip=true]{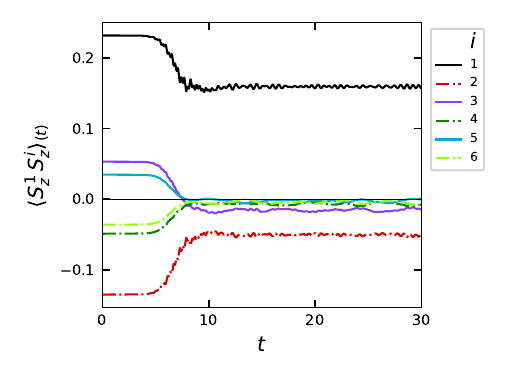}}
		\put(110,80){\includegraphics[height=1.3cm, trim=2.2cm 1.5cm 1.8cm 1.35cm, clip=true]{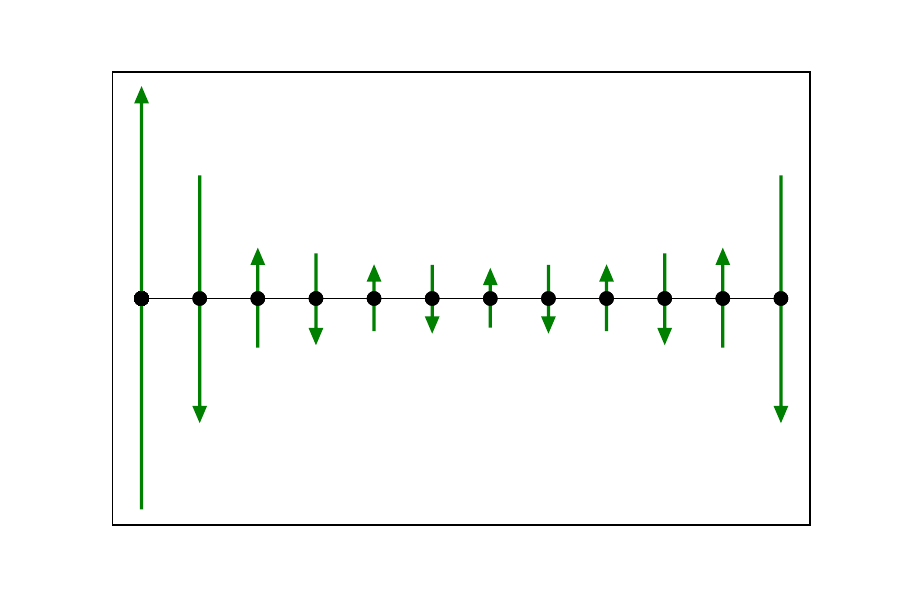}}
		\end{picture}
	}
	\subfloat[$12\!\times\!1$ $v^\prime=0.5$]{
		\begin{picture}(250,170)
		\put(0,0){\includegraphics[width=\TMPWW, trim=0cm 4mm 0cm 0.0cm, clip=true]{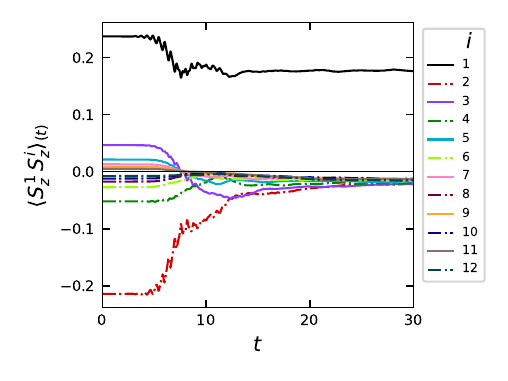}}
		\put(110,93){\includegraphics[height=1.3cm, trim=2.2cm 1.5cm 1.8cm 1.4cm, clip=true]{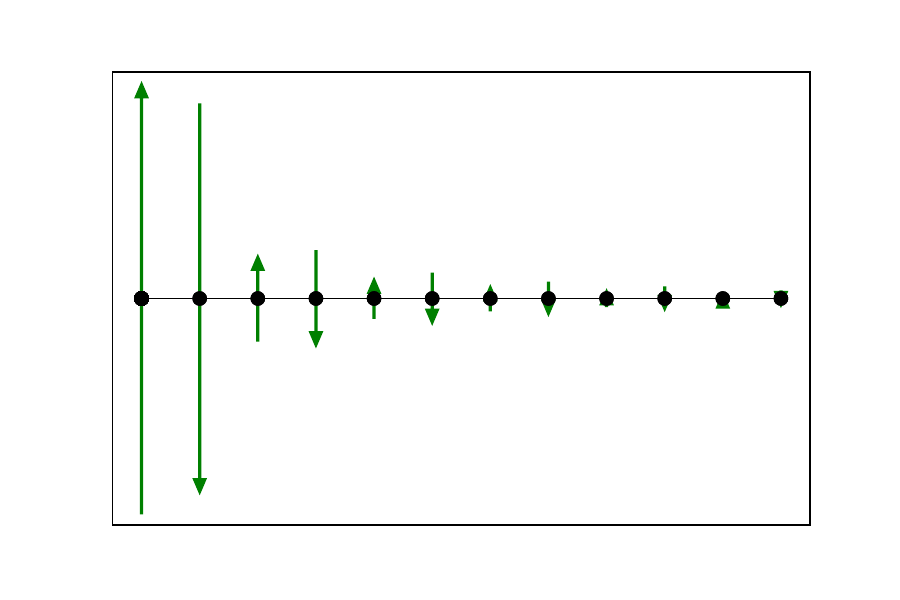}}
		\end{picture}	
	}
	\caption{Equal time spin correlation functions $\braket{\hat S_z^{1,1} \hat S_z^{i,j} }(t)$ (for box geometries) and $\braket{\hat S_z^{1} \hat S_z^{j} }(t)$ (for chains), with the first spin operator kept fixed at the upper left site. In all plots different colors denote different sites for the second spin operator. In case of chains (c)-(e) one index $j$ is enough. For box geometries (a,b) a pair of indices $i$ and $j$ is used to denote on which site the second operator is placed. OBC are used, except for (d). Insets: The values for $t=0$ plotted as green arrows with the length proportional to the magnitude of the correlation function.}
	\label{fig::StaticSpinCorr}
\end{figure*}

\subsubsection{Spin-spin correlation function}

The time-dependent equal time ($t=t'$) spin-spin correlation functions $\braket{\hat S_z^{1} \hat S_z^{j} }(t)$ for the $12$-site clusters are shown in \cref{fig::StaticSpinCorr}.  For the ground states ($t=0$) of all investigated clusters, strong antiferromagnetic correlations are visible over several lattice sites (see also insets of \cref{fig::StaticSpinCorr}). For the $12\!\times\!1$ system with $v'=0.5$, \cref{fig::StaticSpinCorr}(e), the spins are frustrated which leads to a slightly smaller correlation length. After the light pulse the correlation length decreased significantly in all systems. For the $4\!\times\!3$, $6\!\times\!2$, \cref{fig::StaticSpinCorr}(a-b), and $12\!\times\!1$ $v'=0.5$, \cref{fig::StaticSpinCorr}(e), cases no tendency toward spin order survives the pulse. For 12$\times$1 $v'=0$, \cref{fig::StaticSpinCorr}(c), and 12$\times$1 with periodic boundary conditions (PBC), \cref{fig::StaticSpinCorr}(d), systems the spins at neighboring sites are still correlated antiferromagnetically, but far less than before the pulse. 
The light pulse is strong enough to destroy long-range spin correlations in all the considered Hubbard clusters.

\subsubsection{Energy of the spin system}

\begin{figure*}
	\centering
	\subfloat[potential energy related to double occupancy $E_{\mathrm{docc}}$]{
		\includegraphics[width=\TMPWW, 
		trim=0cm 4mm 0cm 0cm, clip=true]{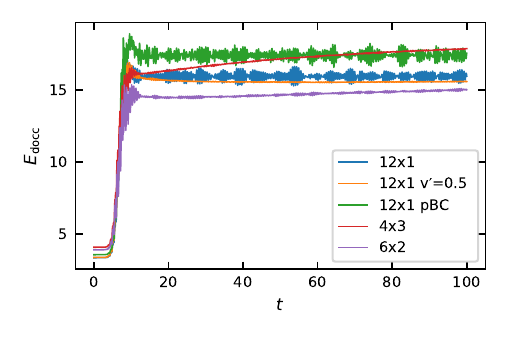}
		\label{fig::Energies_Docc}
	}
	\subfloat[kinetic energy]{
		\includegraphics[width=\TMPWW, 
		trim=0cm 4mm 0cm 0cm, clip=true]{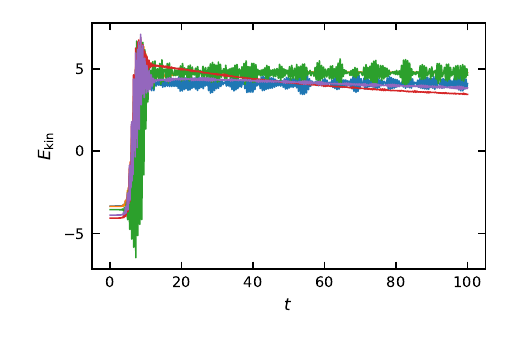}
		\label{fig::Energies_Kin}
	}
	\hspace{0mm}
	\subfloat[Heisenberg spin energy]{
		\includegraphics[width=\TMPWW, 
		trim=0cm 4mm 0cm 0cm, clip=true]{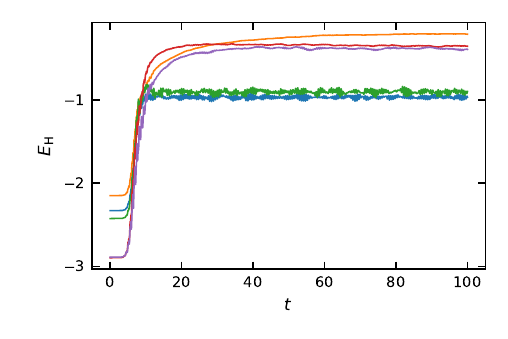}
		\label{fig::Energies_Heisenberg}
	}
	\subfloat[total energy]{
		\includegraphics[width=\TMPWW, 
		trim=0cm 4mm 0cm 0cm, clip=true]{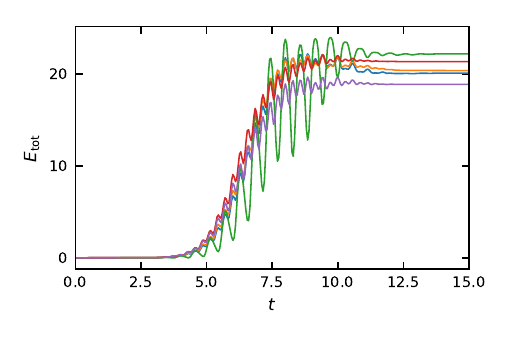}
		\label{fig::Energies_Tot}
	}
	
	\caption{Different contributions to the total energy of $12$-site clusters, as defined in \cref{eqn::EnergyHeisenberg2} and \cref{eqn::EnergiesTotDoccKin}, as a function of time: (a) potential energy related to double occupancy; (b) kinetic energy; (c) Heisenberg spin energy; and (d) total energy. Different colors correspond to different cluster geometries, as denoted in the legend in (a). Please note, that the total energy in (d) does not change after the pulse (centered at $t_p=8$) is over and therefore it is shown only for shorter times. The ground state energy corresponding to $E_\mathrm{tot}(t=0)$ was set to zero. Parameters as described in~\cref{sec::parameters}.}
	\label{fig::Energies}
\end{figure*}

To assess the importance of energy absorbed by the spin system, we show in \cref{fig::Energies_Heisenberg} the Heisenberg spin energy $E_{\mathrm{H}}$ defined in \cref{eqn::EnergyHeisenberg2}. We compare it to the total energy $E_\mathrm{tot}$, the potential energy related to the double occupancy $E_{\mathrm{docc}}$ and the kinetic energy $E_{\mathrm{kin}}$ shown in \cref{fig::Energies} (a-b,d). They are defined as
\begin{equation}
\begin{array}{rcl}
E_{\mathrm{tot}}(t) &=&  \bra{\psi(t)} \hat H(t)  \ket{\psi(t)}  \\ 
E_{\mathrm{docc}}(t) &=& U \sum_i \bra{\psi(t)} \hat n_{i\downarrow}  \hat n_{i\uparrow}  \ket{\psi(t)} \\
E_{\mathrm{kin}}(t) &=&  \sum_\sigma \sum_{i \neq j} v_{ij}(t) \bra{\psi(t)} \hat c^{\dagger}_{j \sigma} \hat c_{i \sigma} \ket{\psi(t)}. 
\end{array}
\label{eqn::EnergiesTotDoccKin}
\end{equation} 
%
%
By definition $E_{\mathrm{tot}}(t)  = E_{\mathrm{kin}}(t) + E_{\mathrm{docc}}(t)$. In \cref{fig::Energies} the total energy at $t=0$ (the ground state energy in our case) was set to zero.

In all systems the additional energy coming from the $E$-field leads to a rise in both kinetic and potential energy. After the pulse is over, the total energy of the systems does not change anymore. We can see that all $12$-site systems considered absorb a similar amount of total energy, see \cref{fig::Energies} (d). In the $4\!\times\!3$ and $6\!\times\!2$ systems the double occupancy, and thus also $E_{\mathrm{docc}}$, rises after the pulse due to impact ionization, which is consistent with \cref{fig::GL_Docc}. The frustrated $12\!\times\!1$ chain with $v'=0.5$ shows a slight decrease of $E_{\mathrm{docc}}$ instead. The other two systems with a $12\times1$ geometry do not show any systematic change in $E_{\mathrm{docc}}$. (The maximal time shown here is $t=100$. For a discussion of recurrence see~\cref{App:fidelity}.)

Comparing the time dependence of $E_{\mathrm{docc}}$ to $E_{\mathrm{H}}$ we do not see the anti-correlation anticipated in Ref.~\onlinecite{Kauch2020Disorder}. Instead, we find an additional rise after the pulse in $E_{\mathrm{H}}$ for the $6\!\times\!2$, $4\!\times\!3$, $12\!\times\!1$  $v'=0.5$, whereas the systems that still remained antiferromagnetically  correlated after the pulse ($12\!\times\!1$ PBC and OBC with  $v'=0$) do not show any systematic change in $E_{\mathrm{H}}$. We also see in \cref{fig::Energies} that $E_{\mathrm{H}}$ and its changes are significantly smaller than the other involved energies. From these considerations, the supposition~\cite{Kauch2020Disorder} that strong spin correlations prevent impact ionization in $12$-site clusters does not seem to be confirmed. The two $12$-site chains with NN-hopping remain, however, distinct in that they absorb less energy into the spin system. This is consistent with \cref{fig::StaticSpinCorr}, where the antiferromagnetic correlations survived the pulse only for these systems.

\subsection{Generalized Loschmidt amplitude for Heisenberg and total energy}
\label{sec:GL_spin}

\begin{figure*}[htb]
	\centering
	\subfloat[$4\!\times\!3$ total]{
		\includegraphics[width=\TMPWW]{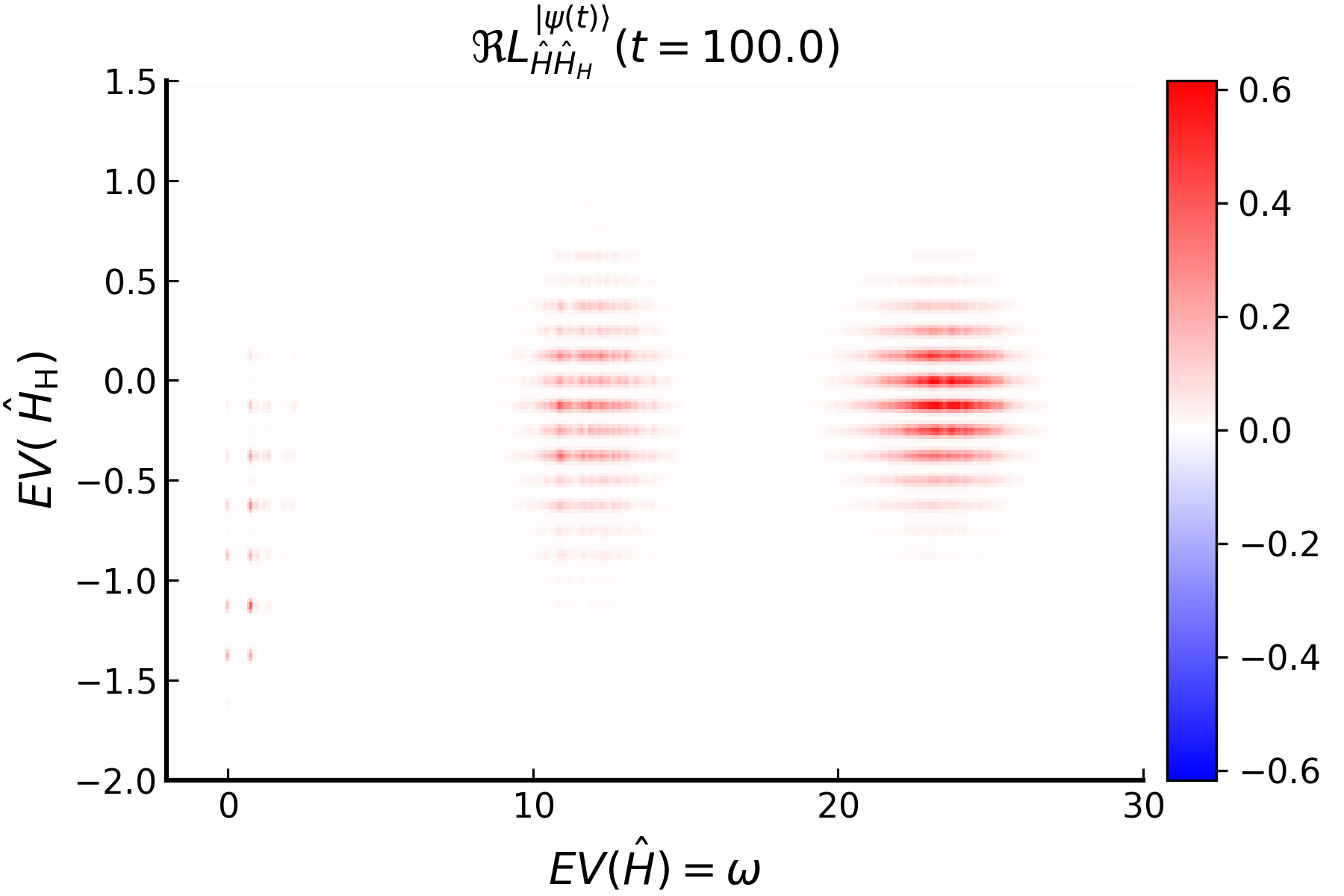}
	}
	\subfloat[$4\!\times\!3$]{
		\includegraphics[width=\TMPWW]{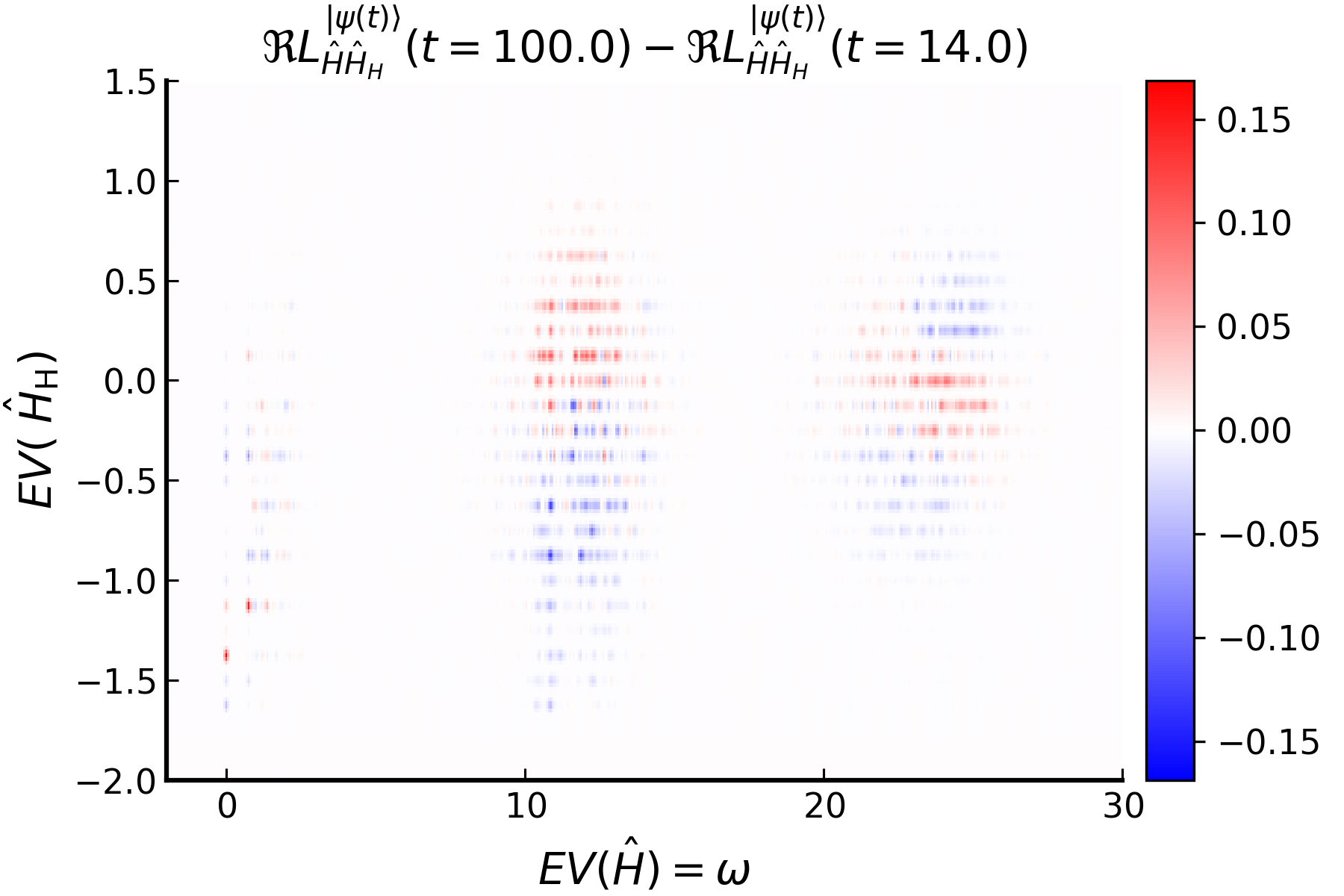}
	}
	\hspace{0mm}
	\subfloat[$6\!\times\!2$]{
		\includegraphics[width=\TMPWW]{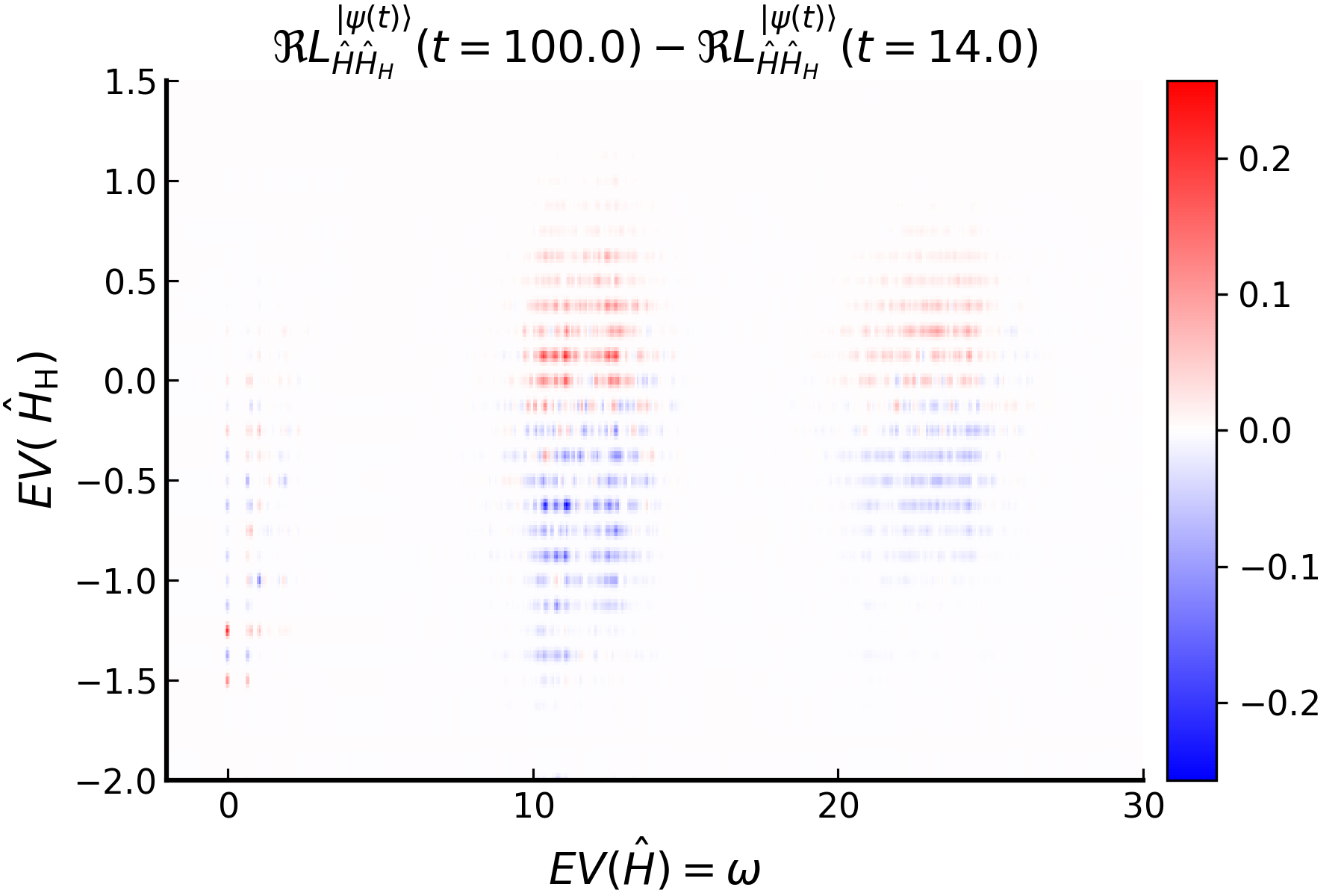}
	}
	\subfloat[$12\!\times\!1$]{
		\includegraphics[width=\TMPWW]{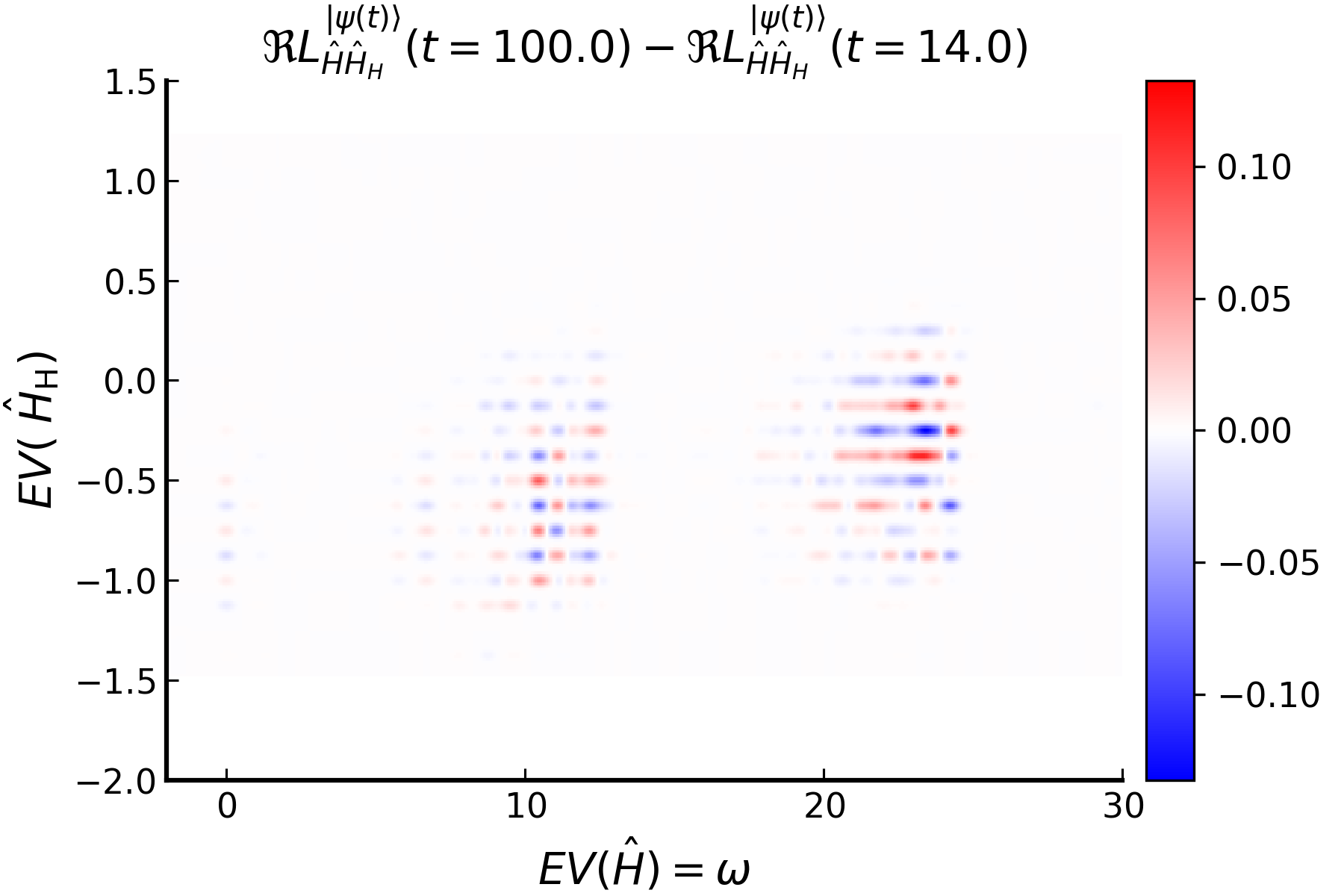}
	}
	\hspace{0mm}
	\subfloat[$12\!\times\!1$ pBC]{
		\includegraphics[width=\TMPWW]{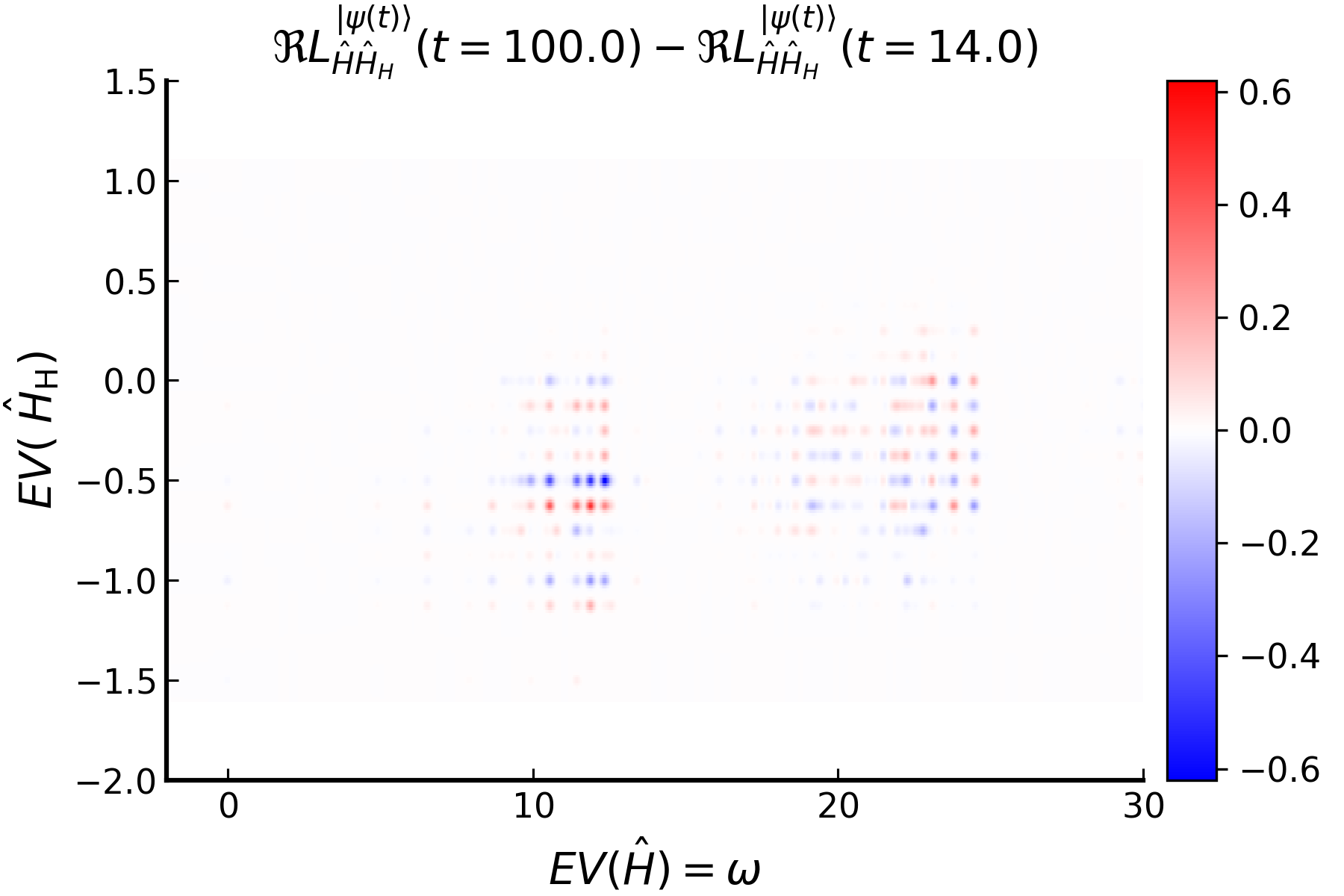}
	}
	\subfloat[$12\!\times\!1$ $v^\prime=0.5$]{
		\includegraphics[width=\TMPWW]{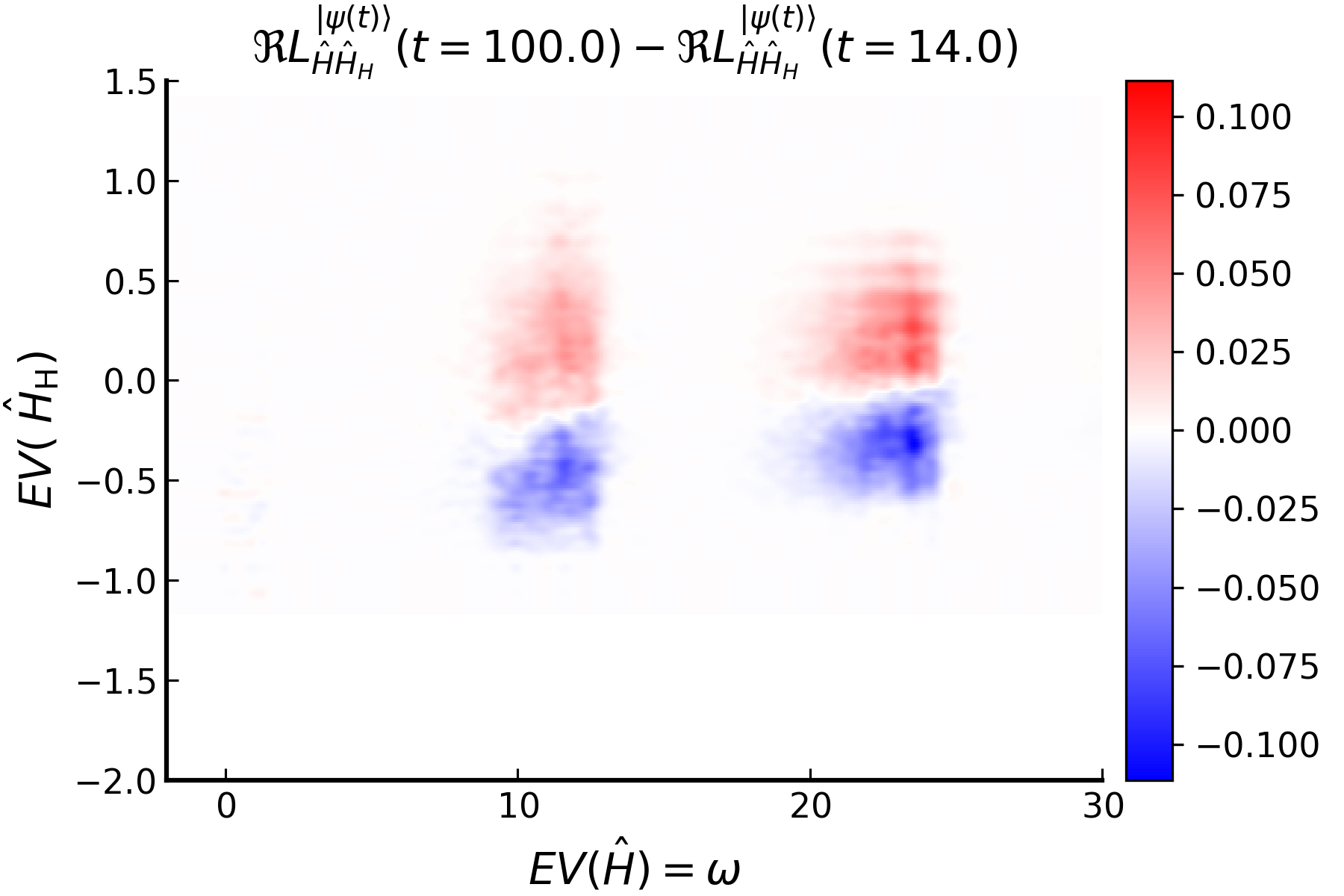}  
	}
	\caption{Generalized Loschmidt amplitude for the $N_s=12$ site systems. In \textit{(a)} the full generalized Loschmidt amplitude is given, in \textit{(b-f)} relative change (differences) after the  the pulse is over are shown. $EV(\hat H)$ broadening is $\sigma_\omega=0.18$; $EV(\hat H_H)$ broadening $\sigma_\omega=0.02$. 
	}
	\label{fig::GL_Spin}
\end{figure*}

We now apply the generalized Loschmidt amplitude as defined in \cref{eqn::LAB_def1} to the spin-correlation energy (Heisenberg energy defined in \cref{eqn::EnergyHeisenberg2}) and total energy, i.e. we take $\hat A = \hat H(0)$ and $\hat B = \hat H_{\mathrm{H}}$.
%
%
In \cref{fig::GL_Spin}(a) we show the real part of the generalized Loschmidt amplitude $L_{\hat{H}\hat H_{\mathrm{H}}}$ for the $4\times3$ system at a time long after the pulse ($t=100$). The system is brought so strongly out of equilibrium that the ground state contribution is almost negligible. The dynamics are dominated by the single- and double-photon excitations. Both give similar eigenvalue contributions for the Heisenberg energy because the spin energy is small compared to the other energy scales (see \cref{fig::Energies}). We observe the same for the other $12$-site clusters (not shown here). In Figs.~\ref{fig::GL_Spin}(b)-(f) we show the difference between   $L_{\hat{H}\hat H_{\mathrm{H}}}$ at $t=100$ and at a shorter time after the pulse $t=14$ for all $12$-site systems considered (analogously as in \cref{fig::GL_Docc}).  

The first system, the $4\times3$ geometery, is  shown in \cref{fig::GL_Spin}(b). We find that the double-photon excitations ($EV(\hat H) \approxeq 2\omega_p$) have almost no effect on the overall expectation value of $E_{\mathrm{H}}$ though there is some internal dynamic. There is a redistribution between the spin-energies $\approxeq 0 $ to larger but also to smaller values which cancel each other. The major states responsible for the long time trend in $E_{\mathrm{H}}$ for the $4\!\times\!3$ system (cf. \cref{fig::Energies_Heisenberg}) are the single-photon ($EV(\hat H) \approxeq \omega_p$) excitations.

The situation is slightly different for the $6\!\times\!2$ and the $12\!\times\!1$ $v'=0.5$ systems (\cref{fig::GL_Spin}(c) and (f), respectively). Here also the double-photon excitations show a clear trend towards further reordering the spins and increasing the Heisenberg spin energy. For these systems, both the single- and the double-photon excitations give important contributions to the long-time behavior of $E_{\mathrm{H}}$. In all three systems, where $E_{\mathrm{H}}$ increases long after the pulse, this growth is also clearly visible in the generalized Loschmidt amplitude $L_{\hat{H}\hat H_{\mathrm{H}}}$ [Figs.~\ref{fig::GL_Spin}(b), (c), and (f)] as an increase in the contribution of states with higher eigenvalues of $\hat{H}_{\mathrm{H}}$.

The situation is very different for the $12\!\times\!1$ systems with PBC and OBC with only NN-hopping ($v'=0$). 
Neighboring contributions regarding EV($\hat H_{\mathrm{H}}$), which differ by a single spin flip, are alternating in sign [see \cref{fig::GL_Docc}(d-e)]. All the dynamics within the same $EV(\hat H)$ (with long time scale) average out. This can also be seen in \cref{fig::Energies_Heisenberg}. The remaining dynamics are on a shorter time scale and thus must be due to larger energy differences, i.e., between the different numbers of photoexcitations. The major contribution comes from the energy difference between the single- (and also double-)photon excitations and the ground state. This explains why the fluctuations in \cref{fig::Energies_Heisenberg} have a frequency of $\omega_p$ or $2\omega_p$ (it is not visible in the figure, but can be extracted from a Fourier transform of the time dependence, which is not shown here).

Looking at Figs.~\ref{fig::GL_Docc} and~\ref{fig::GL_Spin} together, one observes that for three systems, $4\times3$, $6\times2$, and $12\times1$ with $v'=0.5$, there is a clear tendency in the dynamics of the charge and spin excitations after the pulse is over. The contribution of higher eigenvalues  of $\hat{H}_{\mathrm{H}}$ increases (i.e. the spin order is further destroyed) and the double occupancy either increases ( $4\times3$ and $6\times2$) or slightly decreases ($12\times 1$ with $v'=0.5$). In the remaining two $12$-site chains with NN-hopping only, there is no clear tendency in the dynamics, and different contributions to double occupancy and spin energy cancel each other. The residual antiferromagnetic correlation (cf. \cref{fig::StaticSpinCorr}) is not further destroyed and there is also no net increase or decrease in double occupancy.

\section{Conclusions}
\label{sec:Conclusions}

We presented the analysis of the dynamics of small Hubbard clusters during and after photoexcitation with a strong electric pulse, focusing on $12$-site systems with and without impact ionization. To this end, we applied novel commutator-free Magnus integrators~\cite{auzinger2021} for the time evolution (solution of the time-dependent Schr\"odinger equation). 

The eigenenergies where the system can absorb energy (at initial times) can be accurately predicted from the Loschmidt amplitude $L^{\hat j \ket{\psi_0}}$ even for a large electric field. For small fields, our results reduce to Fermi's golden rule. 

On the other hand, the description of optical absorption through the one-particle Green's function turns out to fail for purely one-dimensional systems. Here, vertex corrections play the dominant role. This situation changes when the geometry of the system is changed to a box or a further neighbor hopping is added, which increases the coordination number. There, the one-particle-based 'bubble' contribution already qualitatively resembles the full result and, in particular, reproduces the correct optical gap.


Further, we generalized the Loschmidt amplitude to gain insight which energy states are responsible for the long-time dynamics of the system. Specifically, we applied the generalized Loschmidt amplitude first to the double occupancy and energy. We found that in the $12$-site clusters with strong impact ionization, it originates predominantly from the double-photon excitations. Only in cases where the pulse frequency was made larger than optimal, and consequently the impact ionization was weaker, the single-photon excitations dominate. The dominance of double- and even triple-photon excitations in the dynamics of the double occupation is likely caused by the combination of two factors: the small size of the systems and, at the same time,  the relatively large strength of the electric field used in our study. Our results for weaker fields (not shown in this paper) are inconclusive in this respect because the generalized Loschmidt amplitude fluctuates (relatively) much stronger in time for weaker fields.

The spin dynamics of the $12$-site systems, as reflected in spin-spin correlation functions and in the Heisenberg spin energy, do not confirm the expectations of Ref.~\cite{Kauch2020Disorder}, namely that spin-reordering tendencies might be detrimental to impact ionization. The Heisenberg spin energy increased also for those systems that displayed impact ionization. Furthermore, the energy scales associated with the spin order were shown to be small compared to the other involved energy scales. The generalized Loschmidt amplitude applied to energy and Heisenberg spin energy also showed no clear anticorrelation between creating more double occupations and spin excitations. We conclude that, at least for $12$-site clusters after a strong photoexcitation, spin excitations cannot explain the absence of impact ionization in purely one-dimensional systems (no impact ionization was found in chains as long as $40$-sites, cf. Ref.~\cite{Grabenwarter2020}, where the matrix-product states were used). 

From a more general perspective, we have introduced a new
analysis tool for studying the dynamics out of equilibrium: the generalized Loschmidt amplitude.
We have applied it to study the dynamics of impact ionization in small Hubbard clusters after a light pulse. The generalized Loschmidt amplitude can also be applied to other physical problems and provides a computationally efficient way of getting a detailed picture of the nonequilibrium dynamics.

\vspace*{3mm}
\textit{Acknowledgments.} 
We thank O. Koch and  W. Auzinger for many fruitful discussions.
This work was supported by the Austrian Science Fund (FWF) through project 
P 30819. 
Calculations have been done on the Vienna Scientific Cluster (VSC).

\appendix

\section{Further properties of Loschmidt amplitude}
\label{App:Properties}


The one-operator Loschmidt amplitude can also be expressed in terms of the two-operator Loschmidt amplitude by integration
\begin{equation}
L_{\hat A}^{\ket \psi }(\alpha) = \frac{1}{2 \pi} \int_{- \infty }^\infty \md \beta \: L_{\hat A \hat B}^{\ket \psi }(\alpha, \beta ).
\end{equation}
It moreover fulfills the property
\begin{equation}
\braket{\hat A^n \hat B^m} = \frac{1}{(2 \pi)^2}  \int_{-\infty}^\infty \md \alpha \: 
\md \beta \: \alpha^n  \beta^m L_{\hat A \hat B}(\alpha, \beta).
\label{eqn::AppendixGLProbFoo}
\end{equation}
The analogous property for the 
standard Loschmidt amplitude is:
\begin{equation}
\braket{\hat A^n} = \int_{-\infty}^\infty \md \alpha \: \frac{1}{2 \pi} \alpha^n L_{\hat A}^{\ket{\psi}}( \alpha )
\end{equation}
While the one-operator Loschmidt amplitude is real $L_{\hat A}^{\ket{\psi}}( \alpha ) \in \mathbb{R}$, the two operator variant can have an imaginary part as well (as $[\hat A, \hat B] \neq 0 $ in general). 
For \cref{eqn::AppendixGLProbFoo} only the real part gives a contribution, and the imaginary part vanishes. Therefore, it is sufficient to only consider the real part of the Fourier transformed two-operator Loschmidt amplitude.

\textit{  Expression in terms of projectors --} For some operators, the possible eigenvalues as well as the corresponding projectors $P$ can be computed cheaply without the need to diagonalize a large matrix numerically. For example the double occupancy operator $\hat D = \sum_i \hat n_{\downarrow i} \hat n_{\uparrow i} $ has for a half-filled system with $N_s$ sites the eigenvalues $\{0, 1, ..., N_s/2 \}$.
\footnote{The eigenstates of $\hat D$ can also be computed efficiently by means of bit-shifts in the second-quantization basis.} 
In such a case it is convenient to express (and calculate numerically) the Loschmidt amplitude by means of these projectors (e.g. $\hat B = \sum_b  b \ket b \bra b = \sum_b b \hat P_b$).
A practical way of expressing the Loschmidt amplitude via one operator $\hat A$ where the eigenspectrum is a priori unknown and one operator $\hat B$ where the projectors, as well as the eigenspectrum, are known is given by

\begin{equation}
\begin{array}{rcl}
^{P_b}\!L^{\ket{\psi(t)}}_{\hat A}(\bar \alpha) &=& \phantom{2\pi} \bra{\psi(t)} \hat P_b \, \me^{-\mi \bar \alpha \hat A} \ket{\psi(t)} \\
^{P_b}\!L^{\ket{\psi(t)}}_{\hat A}(\alpha)      &=& 2\pi \sum_a \bra{\psi(t)} \hat P_b \ket{a}  \braket{a | \psi(t)} \\
&& \quad \quad \quad \times \delta(\alpha-a)
\end{array}
\label{eqn::LAB_def3}
\end{equation} 
The set of functions for all $b$ has exactly the same information content as \cref{eqn::LAB_def1} but it it is cheaper to compute. 

\paragraph*{Relation to the optical conductivity}--
The Loschmidt amplitude of $\ket{\psi} = \hat j \ket{\psi_0}$ is related to the retarded current-current correlation function as
\begin{equation}
\begin{array}{rcl}
K^R(t,0) &=& \mi \theta(t) \, \braket{[\hat j(t), \hat j(0)]}_0 \\
&=& - 2 \theta(t) \, \Im L^{\hat j \ket{\psi}_0}(t).
\end{array}
\label{eqn::OptCondTimeAppendix}
\end{equation}
For convenience we also introduce the current-current correlation function as $K(t) = \mi \braket{[\hat j(t), \hat j(0)]}_0$.

In frequencies this amounts to 
\begin{equation}
L^{\hat j \ket{\psi}_0}(\omega > 0) = 2 \: \Im K^{R}(\omega > 0 ).
\label{eqn::OptCondFreqAppendix}
\end{equation}
\cref{eqn::OptCondFreqAppendix} can be derived as follows:
we start by separating the Loschmidt amplitude into a symmetric and an anti-symmetric part. 
\begin{equation}
{L^{\hat j \ket{\psi}_0}}(\omega) \equiv \frac{1}{2} {L^{\hat j \ket{\psi}_0}}^s(\omega) + \frac{1}{2} {L^{\hat j \ket{\psi}_0}}^a(\omega).
\label{eqn::AppProve1_01}
\end{equation}
By definition 
\begin{equation}
\begin{array}{rcl}
{L^{\hat j \ket{\psi}_0}}^s(\omega) &=& \phantom{-}{L^{\hat j \ket{\psi}_0}}^s(-\omega)\\
{L^{\hat j \ket{\psi}_0}}^a(\omega) &=& -{L^{\hat j \ket{\psi}_0}}^a(-\omega).
\end{array}
\label{eqn::AppProve1_01a}
\end{equation}
Because $L^{\hat j \ket{\psi}_0}(\omega) \in \mathbb{R}$ (cf. \cref{eqn::L_FGR2}) for $\omega >0$ and $\omega<0$,   also ${L^{\hat j \ket{\psi}_0}}^s(\omega) \in \mathbb{R}$ and ${L^{\hat j \ket{\psi}_0}}^a(\omega) \in \mathbb{R}$. 
We further use that $L^{\hat j \ket{\psi}_0}(\omega < 0)=0$ (groundstate energy is chosen as zero) implies that the symmetric part must cancel the anti-symmetric part for negative frequencies. Due to their respective (anti-)symmetry it follows that for positive frequencies they both must be equal to the full function
\begin{equation}
{L^{\hat j \ket{\psi}_0}}(\omega > 0) = {L^{\hat j \ket{\psi}_0}}^s(\omega > 0) = {L^{\hat j \ket{\psi}_0}}^a(\omega > 0).
\label{eqn::AppProve1_02}
\end{equation}
The corresponding Fourier transforms are purely real/imaginary. 
As a next step consider 
\begin{equation}
\begin{array}{rcl}
K(t) &=& -2 {L^{\hat j \ket{\psi}_0}}(t)\\
&=& -2 \Im \left( \frac{1}{2} {L^{\hat j \ket{\psi}_0}}^s(t) + \frac{1}{2} {L^{\hat j \ket{\psi}_0}}^a(t)\right) \\
&=& - \Im {L^{\hat j \ket{\psi}_0}}^a(t) \\
&=& \mi {L^{\hat j \ket{\psi}_0}}^a(t).      
\end{array}
\label{eqn::AppProve1_02b}
\end{equation}
The first line follows directly from the definition of $L^{\hat j \ket{\psi}_0}(t)$. In the second line \cref{eqn::AppProve1_01} was inserted. In the third line we used that ${L^{\hat j \ket{\psi}_0}}^s(t)$ is purely real. In the last line we used that ${L^{\hat j \ket{\psi}_0}}^a(t)$ is purely imaginary.  
%
In frequencies \cref{eqn::AppProve1_02b} can also be written as
\begin{equation}
\mi {L^{\hat j \ket{\psi}_0}}^a(\omega) = K(\omega ).
\label{eqn::AppProve1_04}
\end{equation}
The last step is now to relate $K(\omega)$ to $\Im K^R(\omega)$.
This can be done by 
splitting up $K(t)\in \mathbb{R}$ into a symmetric and an anti-symmetric part and using the property that $K(t) = - K(-t)$  is anti-symmetric in time. Thus the anti-symmetric part  $K^{R \: a}(t)$ is given by $K(t)$:
\begin{equation}
\begin{array}{rcl}
K^R(t) &\equiv& \frac{1}{2} K^{R \: s}(t) + \frac{1}{2} K^{R \: a}(t) \\
&=& \frac{1}{2} K^{R \: s}(t) + \frac{1}{2} K(t)
\end{array}
\label{eqn::SymRelationChi1}
\end{equation}
Taking the imaginary part of the Fourier transform of \cref{eqn::SymRelationChi1} (and using $K^{R \: s}(\omega) \in \mathbb{R}$) leads to
\begin{equation}
K(\omega) = 2\mi \Im K^R(\omega)
\label{eqn::SymRelationChi}
\end{equation}
Inserting \cref{eqn::SymRelationChi}  into \cref{eqn::AppProve1_04} yields 
\begin{equation}
{L^{\hat j \ket{\psi}_0}}^a(\omega) = 2\Im K^R(\omega).
\label{eqn::AppProve1_05}
\end{equation}
Thus using \cref{eqn::AppProve1_02} we have proven \cref{eqn::OptCondFreqAppendix}.

\section{Equilibrium Green's function}
\label{App:GF}
\begin{figure}[htbp]
	\centering
	\includegraphics[width=\linewidth]{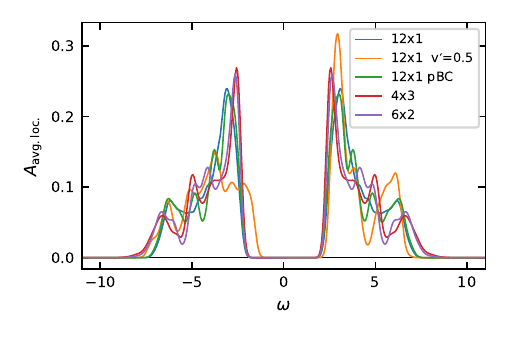}
	\caption{ Local average equilibrium Green's function for all considered $N_s=12$ systems. It is related to the retarded Green's function according to $A_{\mathrm{avg.\, loc.}}(\omega) =  \frac{1}{N_s}\sum_i \frac{-1}{\pi} \Im G^R_{ii}(\omega)$. }
	\label{fig:EqGF}
\end{figure}

In~\cref{fig:EqGF} we show the local spectral function  $A_{\mathrm{avg.\, loc.}}(\omega)$  (averaged over sites) for all the $12$-site systems considered in the main text. Since direct calculation in the Lehmann representation is unfeasible due to memory constraints, $A(\omega)$ was obtained here by time propagation with the time-independent Hamiltonian. Since, for finite systems, the spectral function consists of a set of $\delta$-peaks, we used a Gaussian broadening of $\sigma_\omega=0.13$.

In all systems where we only considered NN hopping, the spectral function is particle-hole symmetric.
This is, however, not the case in the $12$-site chain with NNN hopping of $v'=0.5$ (as clearly seen in the~\cref{fig:EqGF}).    

\section{Optical conductivity in strong-coupling}
\label{app:Lyo}
\begin{figure}[htb]
	\centering
	\includegraphics[width=\linewidth]{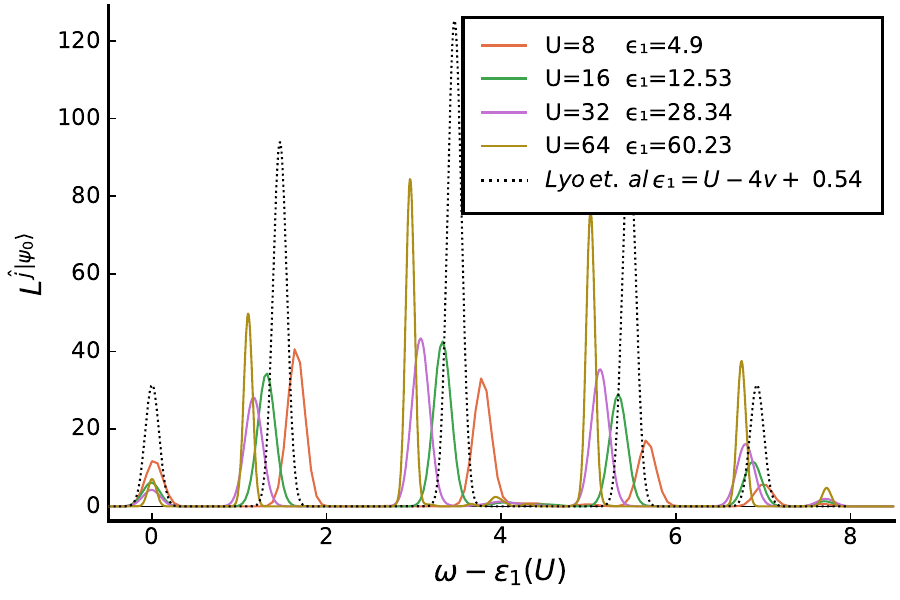}
	\caption{ Loschmidt amplitude  $L^{\hat j \ket{\psi_0}}$ corresponding to $2 \cdot \sigma^R(\omega) \cdot \omega $ for $\omega>0$ for the $12\!\times\!1$ system with PBC. $\epsilon_1$ corresponds to the lowest energy at which the system can absorb light and is always slightly larger than $U-4v$. The analytical results of a large U expansion from Lyo et. al \cite{Lyo1977} are shown as the dotted line. }
	\label{fig:Scaling_12x1pBC}
\end{figure}

\Cref{fig:Scaling_12x1pBC} shows the Loschmidt amplitude/ optical conductivity for $12\times1$ systems with periodic boundary conditions for different interaction values $U$. As a comparison also analytical results of Lyo et al.~\cite{Lyo1977} in the large $U$ limit are shown (dotted line). The peak structures are due to the different (relative) momentum values of the electron-hole pair created on a nearly antiferromagnetic background. In Ref.~\onlinecite{Lyo1977} the background (ground state) considered was antiferromagnetic. In contrast, while being close to the antiferromagnetic phase, we have paramagnetic (non-degenerate) ground states per construction, which is likely the cause of the difference in position of the peaks in our results and the analytic results of Ref.~\onlinecite{Lyo1977}. 


\section{Generalized Loschmidt amplitude for a  \texorpdfstring{$4\!\times\!2$}{4 x 2} system. }
\label{App:OtherFigures4x2}

\begin{figure}[htbp]
	\centering
	\includegraphics[width=\linewidth, trim=0mm 5mm 0mm -5mm]{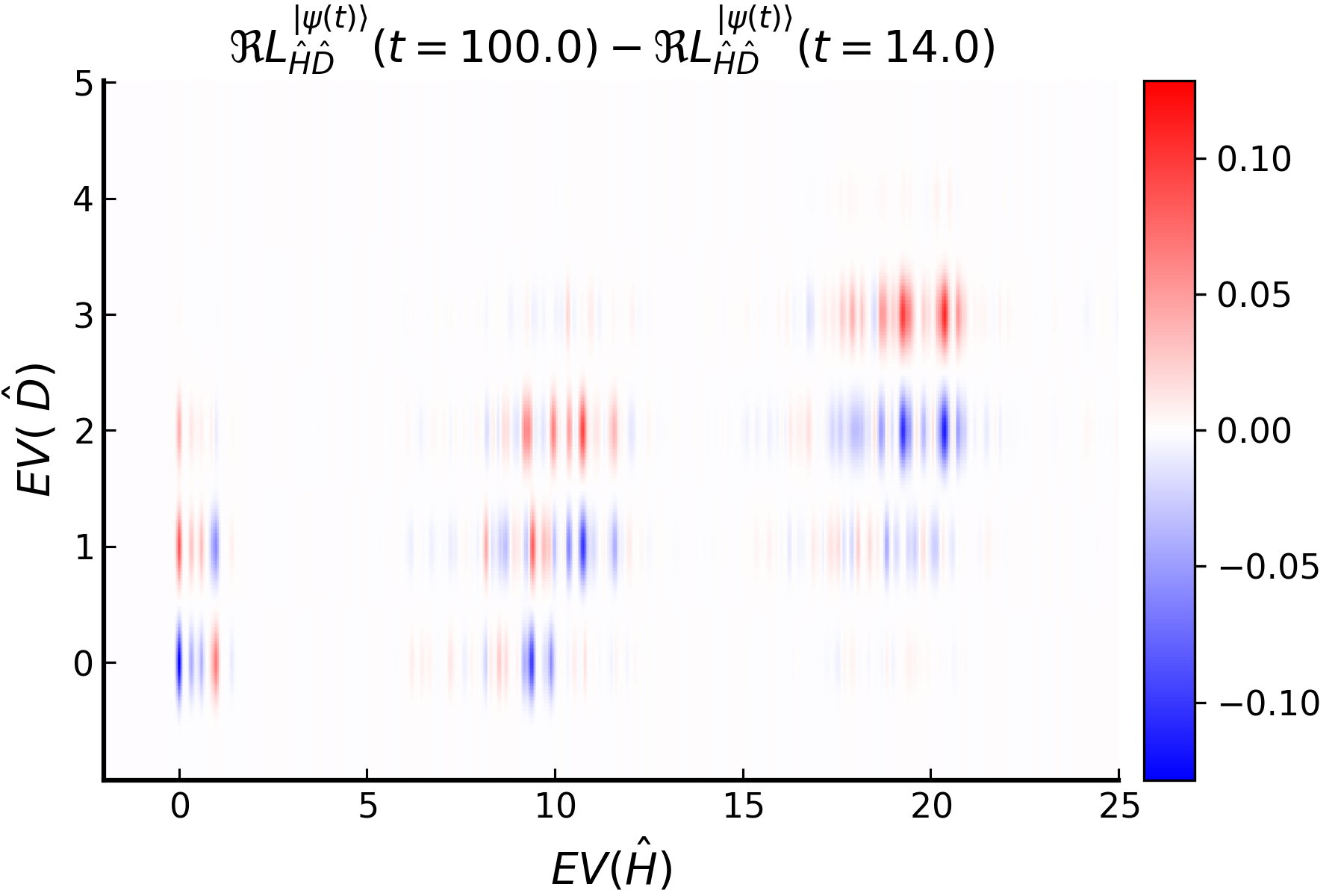}
	\caption{ Generalized Loschmidt amplitude for a $4\!\times\!2$ $v'=0.8$ system. The parameters are as in \cite{Kauch2020Disorder} namely $U=6$  $a=0.8$, $\omega_p=9$. Impact ionization originates in this system also from single photon excitation.}
	\label{fig:GL4x2vd07}
\end{figure}

In~\cref{fig:GL4x2vd07} we show the generalized Loschmidt amplitude defined in~\cref{eqn::LHD_def2} for a $4\times2$ system. In this system significant contributions to  impact ionization stem from single- and double photon-excitations. 


\section{Parameter scan for  $12$-site clusters}
\label{App:Scan}

\begin{figure}[htb]
	\centering
	\includegraphics[width=\linewidth]{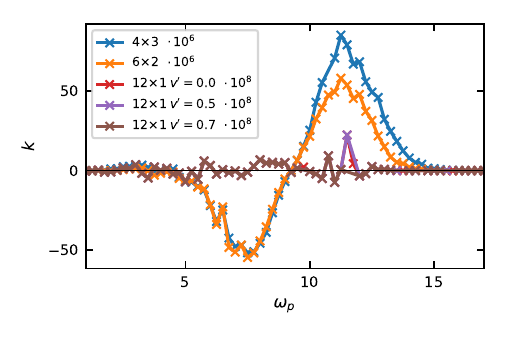}
	\caption{Average rise of double occupation
		according to \cref{eqn:AverageRiseDocc} scaled with $10^6$ or $10^8$ for different geometries. The optimal value of $\omega_p$ for $4\!\times\!3$ and $6\!\times\!2$ is at $\omega_p=11.25$. } 
	\label{fig:parameter_scan}
\end{figure}

As a criterion for impact ionization one can use the average rise of double occupation after the pulse (in other words, $k$ is the average slope of the curve $d(t)$ for times after the pulse is over). We define it in the following way:
\begin{align} 
k &= \frac{\overline{d}_2 - \overline{d}_1}{t_2 - t_1}\label{eqn:AverageRiseDocc}\\ 
\overline{d}_i &= \frac{\int_{t_{i}-T}^{t_{i}+T} \md t\; w(t, \mu=t_i, \sigma)\,  \braket{\hat d(t)}}{\int_{t_{i}-T}^{t_{i}+T} \md t\; w(t, \mu=t_i, \sigma) } \\ 
w(t, \mu, \sigma) &= \me^{\frac{-(t-\mu)^2}{2 \sigma^2}},  
\end{align}
where the times $t_1$ and $T$ need to be chosen such that the Hamiltonian is, within a good approximation, already  time-independent. We chose $t_1-T=t_p + 3\sigma_p$ with $\sigma_p = 2, T=9, t_p=8, t_1=23$. For $t_2$ a time long after the pulse $t_2=300$ was used. The weight-function makes the average rise quite independent to small changes in $t_1$ and $t_2$. In the limit $\sigma\rightarrow\infty$ a simple uniform average is retained. For a rather fast fluctuating function $\sigma \approx \mathcal{O}(T/2)$ seems a reasonable choice and we chose $\sigma=3$ for the results presented in Fig.~\ref{fig:parameter_scan}. 

For all calculations presented in this work, we used the E-field strength parameter $a=0.8$. The effect of field strength on impact ionization was analyzed in \cite{Maislinger2020} (for instance in Fig. 6 there) for the $4\times 3$ cluster. 



\section{Spin susceptibility}
\label{App:Spin}

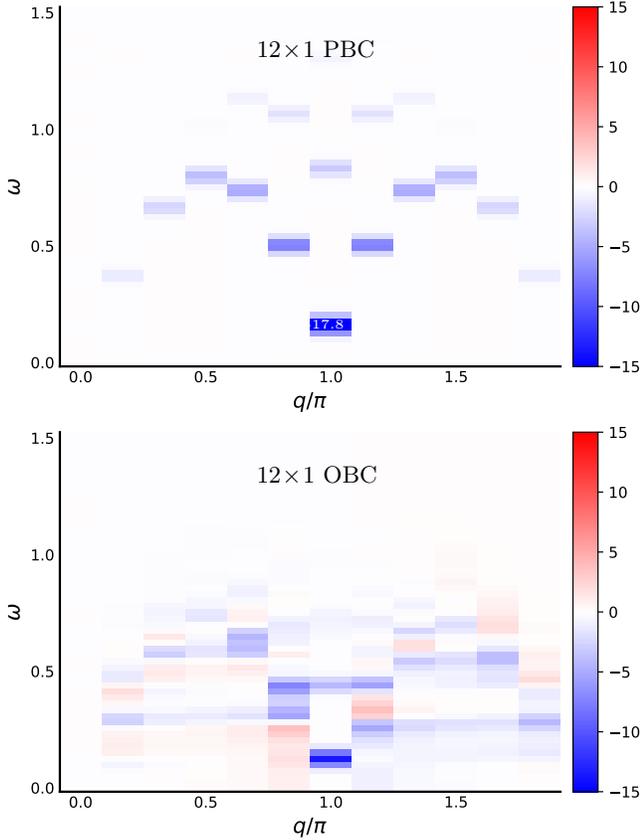
\begin{figure} 
	\centering
	\begin{picture}(255,150)
	\put(0,0){\includegraphics[width=\linewidth]{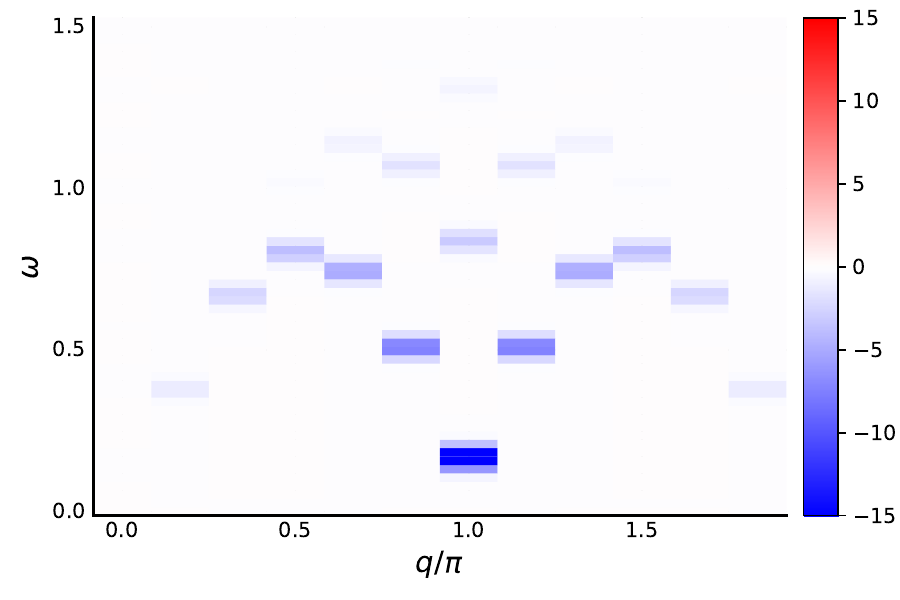}}
	\put(100,140){$12\!\times\!1$ PBC}
	\put(118.5,37.5){\color{white}{\tiny{-17.8}}}
	\end{picture}
	\begin{picture}(255,150)
	\put(0,-10){\includegraphics[width=\linewidth]{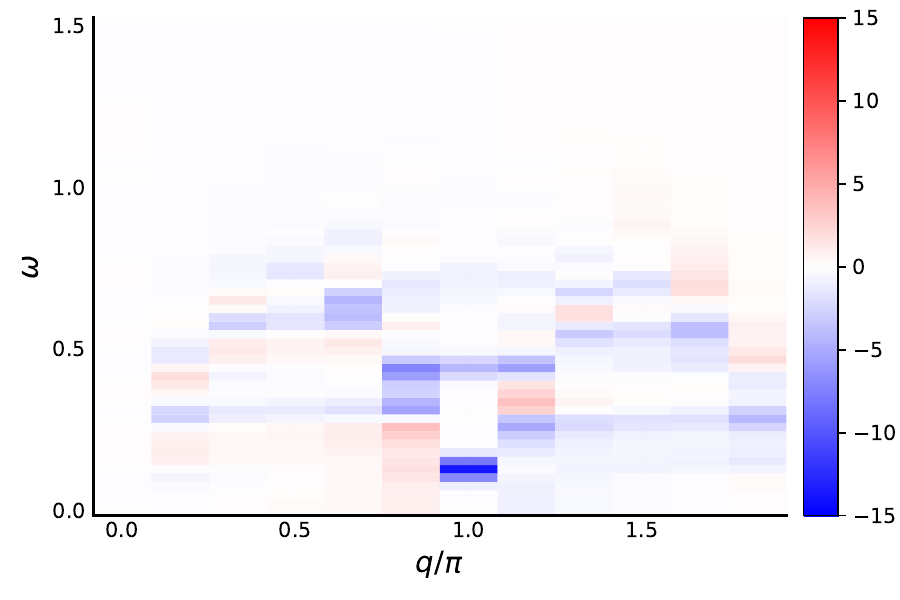}}
	\put(100,130){$12\!\times\!1$ OBC}
	\end{picture}
	\vspace{0mm}
	\caption{ Spin susceptibility $\Im \chi^R(q, \omega)$ for a $12\!\times\!1$ system with PBC (top) and OBC (bottom). For PBC the maximal absolute value $ \max |\Im \chi(q=\pi, \omega)|=17.8$ is outside the colorbar and instead labeled explicitly  to increase the visibility of the dispersion relation. }
	\label{fig:DispersionRelation}
\end{figure}

We  consider the dynamic  spin susceptibility in equilibrium. In \cref{fig:DispersionRelation}, we show $\Im \chi^R(q, \omega)$ for the $12\!\times\!1$ systems with PBC (upper panel) and OBC (lower panel). The double-wing structure of the paramagnon dispersion relation $\propto \left|\sin(q)\right|$ is clearly visible~\cite{Blundell2001magnetism}.  The gap at $q=\pi$ is system-size dependent. It is equal to $J$ for a Hubbard dimer and vanishes for an infinitely large system. At $q=0$ the imaginary part of the susceptibility is exactly zero because the total spin in the system is a conserved quantity. For OBC the momentum $q$ is no longer a good quantum number and  $\Im \chi(q, \omega)$ changes sign~\footnote{For PBC $\chi^R(q_1, q_2, \omega) := \sum_{R_i, R_j} \int_{-\infty}^{\infty} \md t \, \me^{-\mi \omega t} \, \me^{-\mi q_1 R_i}  \, \me^{-\mi q_2 R_j} \times \chi^R_{R_i, R_j}(t)\propto \delta_{q_1, -q_2}$. For OBC this is not the case. In \cref{fig:DispersionRelation} the diagonal parts $\chi^R(q_1,\omega) := \chi^R(q_1,-q_1,\omega)$ are shown.}. Nonetheless, also for this case one can still find a resemblance to the paramagnon dispersion relation similar to the case of PBC. 
The spin susceptibility is still quite large, but smaller than for the PBC. In both cases, the 
$\omega$-q dispersion has an amplitude  $\sim J=4 v_{ij}^2/U=0.5$, and the finite-size spin gap is small enough ($\sim 0.1$) for excess kinetic energy to be transferred to the spin system.
%

\section{Recurrence time}
\label{App:fidelity}

\begin{figure}[htb]
	\centering 
	\includegraphics[width=\linewidth]{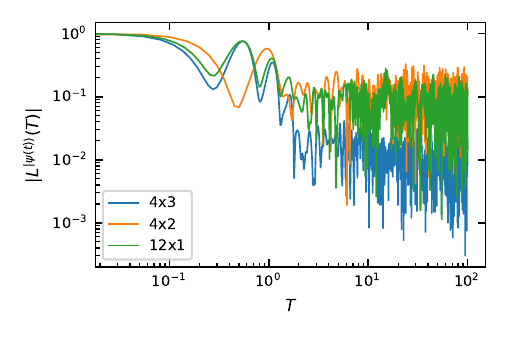}
	\caption{Absolute value of the Loschmidt amplitude. Parameters as in Fig.~\ref{fig::GL_Docc}. At $T=100$ no recurrence has occurred. } 
	\label{fig:Fidelity1} 
\end{figure}

For finite isolated systems with a time-independent Hamiltonian it is expected that after a finite time $T_{\mathrm{rec.}}$ \textit{all} observables have the same values again in the sense that 
\begin{equation}
\braket{\hat{\mathcal{O}}_i(t+T_{\mathrm{rec.}})} \approx \braket{\hat{\mathcal{O}}_i(t)}. 
\label{eqn::RecursionCriterionStrong}
\end{equation}
For a true recurrence to happen (``$=$'' instead of ``$\approx$'' in the equation above) one would need \textit{all} differences between occupied eigenstates that are coupled by the operators $\hat{\mathcal{O}_i}$ to be an integer multiple of some energy $E_{n1} - E_{n2} = m \epsilon_{\mathrm{rec.}}$ with $m \in \mathbb{Z}$. After a time $T_{\mathrm{rec.}}= 2\pi/\epsilon_{\mathrm{rec.}}$ the recurrence criterion would be fulfilled. 

An exact evaluation for a reasonably large system is not possible in a straightforward way due to memory constraints. A lower bound might, however, be given by the smallest energy difference between eigenstates ($m=1$). For instance, for the $4\times2$ system where an exact diagonalization to obtain all eigenenergies is still possible, this naive estimate gives $\epsilon_{\mathrm{rec.}}=2\cdot 10^{-7}$. For a larger system the range of the many-body eigenenergies is expected to grow linear with system size. The dimension of the Hilbert space, on the other hand, will grow exponentially. Thus the eigenenergy differences and consequently also the smallest one of them is expected to decrease exponentially with system size. 

The criterion for a true recursion in \cref{eqn::RecursionCriterionStrong} is too strong to be fulfilled by all finite systems with a time-independent Hamiltonian\footnote{A simple counter example is given by a three-level system with energy levels at $0$, $1$ and $\sqrt{2}$. A true recursion cannot happen in such a system, but \cref{eqn::RecursionCriterionWeak} can be fulfilled for an arbitrarily small $\delta_1$.}.  
We might weaken it and then arrive at the following statement  \cite{Boccieri1957}:
For any finite closed system described by a time-independent Hamiltonian the wave-function $\ket{\psi(t)}$ returns arbitrarily close to itself after a finite time $T_{\mathrm{rec.}}$ such that 
\begin{equation}
\norm{\:\ket{\psi_{(t+T_{\mathrm{rec.}})}} -  \ket{\psi_{(t)}}\:}_2 < \delta_1, 
\label{eqn::RecursionCriterionWeak}
\end{equation}
where the recursion time will depend on the demanded degree of closeness $\delta_1$. 
We could estimate it in the following way. For a recursion in this sense to happen the smallest integers $m_n$ need to be found that satisfy $|T_{\mathrm{rec.}} E_n - 2\pi m_n |<\delta_2$. This is a simple analog of the Poincar\'e recurrence time for a classical system.

A measure for the recurrence of a system can be given by \cref{eqn:ReccurenaceFidelity} which is related to the fidelity of the system \cite{Gorin2006}: 



\begin{equation}
\mathcal{F}(t) = |L^{\ket{\psi}_{(t)}}_{(T)}| = |\langle \psi_{(t)}| \psi_{(t + T)}\rangle|.
\label{eqn:ReccurenaceFidelity}
\end{equation}

\vspace*{0mm}

\noindent In \cref{fig:Fidelity1} the absolute value of the Loschmidt amplitude for a number of different Hubbard clusters is shown. No recurrence was reached for times up to $T=100$. Our simple estimate above would suggest that $T_{\mathrm{rec.}}>10^7$.



%

\end{document}